\documentclass[reprint,superscriptaddress,amsmath,amssymb,aps,pra]{revtex4-2}

\usepackage{graphicx}
\usepackage{svg}
\usepackage{graphics}
\usepackage{booktabs}
\usepackage{dcolumn}   
\usepackage{bm}        
\usepackage{hyperref}  
\usepackage{mathtools}
\usepackage{appendix}
\usepackage{makecell}
\usepackage{orcidlink}
\DeclareMathOperator*{\argmin}{arg\,min} 

\usepackage{xcolor}

\begin{document}
\newcommand {\urss}[1]{\ensuremath{_{\mathrm{#1}}}}
\newcommand {\td}[1]{\protect \todo{#1}} 
\newcommand {\NIFS}{\affiliation{National Institute for Fusion Science, Toki, Gifu 509-5292, Japan}}
\newcommand {\UMD}{\affiliation{Department of Astronomy, University of Maryland, College Park, MD 20742}}
\newcommand {\NASA}{\affiliation{Astrophysics Science Division, NASA/GSFC, Greenbelt, MD 20771}}
\newcommand {\CRESST}{\affiliation{Center for Research and Exploration in Space Science and Technology, NASA/GSFC, Greenbelt, MD 20771}}
\newcommand {\UEC}{\affiliation{Institute for Laser Science, The University of Electro-Communications, Tokyo 182-8585, Japan}}
\newcommand {\clemson}{\affiliation{Department of Physics and Astronomy, Clemson University, Clemson, SC 29634}}
\newcommand {\cea}{\affiliation{Univ.\ Grenoble Alpes, CEA, CNRS, Grenoble INP, IRIG, SyMMES, 38000 Grenoble, France}}
\newcommand {\UD}{\affiliation{Department of Experimental Physics, University of Debrecen, Hungary, 4026}}
\newcommand {\mi}{\affiliation{Massachusetts Institute of Technology, Cambridge, Massachusetts 02139}}
\newcommand {\msu}{\affiliation{Facility for Rare Isotope Beams and Department of Physics and Astronomy,
Michigan State University, East Lansing, Michigan 48824}}
\newcommand {\nur}{\affiliation{Institut für Theoretische Physik II, Universität Erlangen-Nürnberg, Staudtstrasse 7, D-91058 Erlangen, Germany}}
\newcommand{\KU}{\affiliation{Interdisciplinary Graduate School of Engineering Sciences, Kyushu University, Fukuoka 816-8580, Japan}}

\preprint{APS/123-QED}

\title{Extreme Ultraviolet Spectroscopy of Highly Charged Lu and Yb Ions\\for Nuclear Charge Radius Determination}


\author{H. Staiger\,\orcidlink{0009-0000-6856-0037}}
\clemson
\email[Contacts: ]{etakacs@clemson.edu, hstaige@clemson.edu, \\kimura.naoki@nifs.ac.jp, n\_nakamu@ils.uec.ac.jp}

\author{E. Takacs\,\orcidlink{0000-0002-2427-5362}}
\clemson
\NIFS

\author{S.A. Blundell\,\orcidlink{0000-0002-5260-002X}}
\cea

\author{N. Kimura\,\orcidlink{0000-0002-4088-239X}}
\NIFS

\author{H.A. Sakaue\,\orcidlink{0000-0003-2209-3255}}
\NIFS

\author{R.F. Garcia Ruiz\,\orcidlink{0000-0002-2926-5569}} 
\mi

\author{W. Nazarewicz\,\orcidlink{0000-0002-8084-7425}}
\msu

\author{P.-G. Reinhard\,\orcidlink{0000-0002-4505-1552}}
\nur

\author{C.A. Faiyaz}
\clemson

\author{C. Suzuki\,\orcidlink{0000-0001-6536-9034}}
\NIFS

\author{Dipti}
\clemson

\author{I. Angeli}
\UD

\author{Yu.~Ralchenko\,\orcidlink{0000-0003-0083-9554}}
\UMD
\NASA
\CRESST

\author{I. Murakami\,\orcidlink{0000-0001-7544-1773}}
\NIFS

\author{D. Kato\,\orcidlink{0000-0002-5302-073X}}
\NIFS
\KU

\author{Y. Nagai}
\UEC

\author{R. Takaoka}
\UEC

\author{Y. Miya\,\orcidlink{0009-0001-2034-5722}}
\UEC

\author{N. Nakamura\,\orcidlink{0000-0002-7009-0799}}
\UEC

\begin{abstract}
We report a high-precision determination of the natural-abundance-averaged nuclear charge-radius difference between Yb and Lu using extreme ultraviolet (EUV) spectroscopy of highly charged ions (HCIs). By measuring the $D_1$ transition energies in Na- and Mg-like charge states of Lu and Yb confined in the Tokyo electron-beam ion trap, we extract meV-level energy shifts that are directly sensitive to nuclear-size effects. Transition-energy differences obtained from these spectra are compared with state-of-the-art relativistic many-body perturbation theory, including a new treatment of Mg-like ions. We develop a generalized framework to propagate uncertainties arising from nuclear deformation and surface diffuseness and evaluate corresponding nuclear-sensitivity coefficients. Combining Na- and Mg-like results yields mutually consistent radius differences, demonstrating the robustness of both the experimental calibration and the theoretical predictions. To determine absolute isotopic radii, we perform a generalized least-squares optimization incorporating our HCI constraints together with optical-isotope-shift data and muonic-atom results. This analysis establishes that the $^{175}$Lu charge radius is smaller than that of $^{174}$Yb, restoring the expected odd–even staggering across the $N=94$ isotonic chain. Our recommended value, $R(^{175}\text{Lu}) = 5.291(11)$~fm, reduces the uncertainty of the Lu radius by a factor of three compared with the previous electron-scattering result and resolves a long-standing anomaly in rare-earth nuclear systematics. This work demonstrates that EUV spectroscopy of HCIs provides a powerful and broadly applicable method for precision nuclear-structure studies in heavy, deformed nuclei. The techniques developed here enable future investigations of isotonic and isoelectronic sequences, including radioactive nuclides and higher-$Z$ systems.
\end{abstract}

\maketitle

\section{Introduction}

Extreme ultraviolet (EUV) spectroscopy of highly charged ions (HCIs) has emerged as a unique tool for precision nuclear structure studies. By probing $3s - 3p$ transitions in Na-like and Mg-like ions, this technique provides direct sensitivity to nuclear size effects, enabling high-precision measurements of nuclear charge radius differences~\cite{silwal_measuring_2018, silwal_determination_2020}. Unlike traditional isotope shift spectroscopy, EUV spectroscopy allows for inter-element comparisons ~\cite{hosier_absolute_2024, Hosier2025} allowing for novel studies of nuclear structure in the high atomic number $Z$ region of the nuclear chart.

In a companion paper~\cite{companionTakacs2025}, we reported a high precision determination of the nuclear charge radius difference between naturally abundant lutetium and ytterbium, restoring the expected pattern of the odd-even staggering of the nuclear charge radius (OESR) along isotonic chains that involve Lu. The result resolves a long-standing anomaly in nuclear systematics and provides a new benchmark for nuclear structure   calculations
~\cite{companionTakacs2025}.

This work details the experimental and theoretical framework underlying that result. We describe the use of the Tokyo electron beam ion trap (EBIT), optimized for EUV spectroscopy, to produce and confine HCIs under ultra-high vacuum and cryogenic conditions~\cite{Nakamura1997}. Lutetium and ytterbium atoms were injected and their spectra were recorded with sub-0.01~nm resolution. Our data analysis pipeline includes image correction, spectral fitting, and a time-dependent wavelength calibration model that accounts for potential drifts in the EBIT's electron beam position. 

Transition energy shifts between Lu and Yb were extracted for both Na-like and Mg-like charge states, and compared to theoretical predictions from our relativistic many-body perturbation theory (RMBPT) calculations. Nuclear sensitivity coefficients were theoretically determined to relate the observed shifts to differences in nuclear size, deformation, and surface diffuseness.

We present a generalized least-squares optimization procedure that incorporates isotope-shift data and theoretical uncertainties to extract the $^{175}\mathrm{Lu}$--$^{174}\mathrm{Yb}$ charge-radius difference. Combined with the accepted muonic-atom value for $^{174}\mathrm{Yb}$, our result yields an absolute charge radius for $^{175}\mathrm{Lu}$.

To interpret the measured Lu–Yb radius difference, we performed nuclear-structure density functional theory (DFT) calculations using Skyrme~\cite{Bender2003} and Fayans ~\cite{Fayans1998,Fayans2000} energy density functionals (EDF) as described in our companion paper~\cite{companionTakacs2025}. These calculations provide predictions for deformation, pairing, and global radii trends in the rare-earth region, and allow comparison with the experimentally observed odd–even staggering. While the models capture the overall evolution of nuclear size along the isotonic chain, they underestimate the magnitude of OESR along this chain, the need for additional precision data in this region.

These results establish EUV spectroscopy of HCIs as a precision technique for nuclear structure studies in heavy, deformed systems. The approach is broadly applicable to future investigations of isotonic and isotopic sequences across the heavy nuclei region of nuclear chart.

\section{Previous Absolute Ytterbium and Lutetium Nuclear Charge Radii Measurements}
\label{sec:prev_meas}

Absolute radius data are scarce for the lutetium isotopes. Aside from Sasanuma’s 1979 Ph.D. thesis~\cite{Sasanuma1979}, which reported a radius of 5.37(3)~fm for $^{175}$Lu, the only other known measurement is from Suzuki’s 1968 Ph.D. thesis~\cite{Su68}, which employed muonic atom spectroscopy. That study reported Fermi distribution parameters ($c = 6.25(10)\,\mathrm{fm}$, $t = 2.07(30)\,\mathrm{fm}$) but did not provide a direct value for the nuclear charge radius. Although Suzuki’s result was included in early muonic atom compilations such as that of Engfer \textit{et al.}~\cite{ESV74}, it saw limited use in later charge radius evaluations due to the adoption of a spherical charge distribution for the deformed Lu nucleus. Consequently, Sasanuma’s thesis result remained the primary reference point for radii in the Lu isotopic chain \cite{Angeli13, fricke_nuclear_2004}.

The most precise values for the stable Yb isotopes come from a muonic atom spectroscopy experiment by Zehnder \textit{et al.} in 1975~\cite{zehnder_charge_1975}. Although the original publication did not report uncertainties for the root-mean-square (\textit{rms}) radii, later evaluations estimated uncertainties of 0.006\,fm \cite{Angeli1999} and 0.002\,fm~\cite{fricke_nuclear_2004}. Additional muonic atom studies include the Ph.D. theses of Adler (1975)~\cite{adler_study_1975} and Bernhardt (1992)~\cite{bernhardt_1992}. These studies reported similar radii to each other and Zehnder's work, with the primary uncertainties stemming from uncertainties in the nuclear polarization calculations.

Several elastic electron scattering studies were also conducted using Yb. The earliest was by Cooper \textit{et al.} in 1976~\cite{cooper_shapes_1976}, which measured $^{176}$Yb, followed by Creswell’s 1977 Ph.D. thesis~\cite{creswell_electron_1977} on the same isotope. While Cooper \textit{et al.} unfortunately did not extract a charge radius, Creswell reported a charge radius of 5.443~fm. These works employed a deformed Fermi distribution rather than a model-independent parameterization such as the Fourier-Bessel expansion~\cite{devries_nuclear_1987}. In contrast, Sasanuma’s 1979 Ph.D. thesis measured the radii of both $^{174}$Yb and $^{175}$Lu to nominal uncertainties of 0.03~fm using a Fourier–Bessel expansion~\cite{Sasanuma1979}. Sasanuma’s radius of 5.41(3)~fm is significantly higher than the 5.308(6)~fm reported by Zehnder \textit{et al.} from muonic atom spectroscopy for $^{174}$Yb~\cite{zehnder_charge_1975}.

As a result, recent evaluations of nuclear charge radii have relied primarily on the muonic atom measurement by Zehnder \textit{et al.} to anchor the Yb isotopes, and on Sasanuma’s electron scattering thesis for the Lu isotopes \cite{Angeli13, fricke_nuclear_2004}. This combination led to an apparent inversion of the expected odd-even staggering across the isotonic Yb--Lu--Hf triplets: the charge radius of the odd-$Z$ Lu isotope appeared significantly larger than those of its even-$Z$ neighbors---opposite to the systematics observed throughout the rare earth region~\cite{Angeli13, fricke_nuclear_2004}. Our new determination of the Yb--Lu radius difference reduces its uncertainty substantially, allows for a new determination of the Lu radius, and resolves a discrepancy in OESR from the previous critical evaluations of charge radii.

\section{Experimental Details}
\subsection{Tokyo EBIT}

The measurements reported in this work were performed using the Tokyo EBIT at the University of Electro-Communications in Tokyo~\cite{Currell1996,Nakamura1997}. The EBIT consists of three main components: an electron gun, a cryogenic trap region with drift tubes, and an electron collector. During the present experiment, the EBIT was operated at an electron beam energy of \(10~\mathrm{keV}\) and an electron beam current of \(100~\mathrm{mA}\). The electron beam was compressed by a magnetic field of \(3~\mathrm{T}\), generated by a superconducting split coil, which focused the electrons into a narrow radius to achieve a high current density for efficient ionization.

The compressed, high-density electron beam ionizes elements injected through a side port, and the resulting ions are confined in the trapping region, which consists of a series of cylindrical drift tube electrodes. Ion trapping is achieved by the combination of a potential well applied to the drift tubes and the space-charge potential of the compressed electron beam. The trapped ions are further ionized through successive electron impact. For the present measurements, a potential well of \(300~\mathrm{V}\) was applied.

To alternately introduce Lu and Yb atoms, two Knudsen cells~\cite{Yamada2006, Nakamura2000} were mounted on the side ports of the Tokyo EBIT. Each cell was aligned with a side window of the drift tube assembly, and the atomic beam was collimated through a pair of 1~mm pinholes to ensure ballistic injection into the trap region. This method avoids the complexity of pulsed injection and is particularly effective for metallic elements with moderate vapor pressures. The operating temperatures were 1000~$^\circ$C for Lu and 350~$^\circ$C for Yb.

A stainless steel pipe with a stop valve was attached to the side of the Knudsen cell for ytterbium injection. This pipe was used to introduce W(CO)$_6$ and Ne gases for observing the spectra of highly charged W and Ne ions for wavelength calibration \cite{kramida_nevii_2006, kramida_neviii_2006, Ralchenko2008, Gillaspy2009}. Ne was also introduced during the Lu and W measurements, as it acted as a source of light, low-charge state ions to evaporatively cool the injected heavy elements. This cooling effect enhanced the population of Na-like and Mg-like charge states, thereby increasing the signal rate and improving the quality of the spectroscopic data. To observe Ne spectra, the trapped ions were dumped at a frequency of 5~Hz to prevent the accumulation of heavier elements, whereas no dumping was applied during the observations of Lu, Yb, and W.

\subsection{EUV Spectroscopy}

Extreme ultraviolet spectra of the trapped ions were recorded using a flat-field grazing incidence spectrometer~\cite{Ohashi_2022}. The spectrometer employs a variable-line-spacing grating optimized for the 5–20~nm wavelength range and is equipped with a Peltier-cooled CCD detector for high-resolution, low-noise imaging. The system is capable of operating in both first- and second-order diffraction modes, enabling flexible trade-offs between spectral coverage and resolution.

The spectrometer is mounted on one of the radial observation ports of the Tokyo EBIT, with a direct line of sight to the trap center without any additional filters used to record spectra. The electron beam itself serves as the effective entrance slit, and the spectrometer’s geometry ensures that the spatial extent of the ion cloud does not degrade the spectral resolution. The spectral resolution achieved in the present measurements was approximately 0.01~nm, consistent with the instrumental width reported in previous EBIT studies~\cite{Koike2022}. This resolution was sufficient to resolve the Na-like and Mg-like $D_1$ (\(3p\,^2P_{1/2} \rightarrow 3s\,^2S_{1/2}\) and \(3s3p\,^3P_{1}\rightarrow 3s^2\,^1S_{0}\)) transitions of interest. 

\begin{figure}
    \centering
    \includegraphics[width=0.975\linewidth]{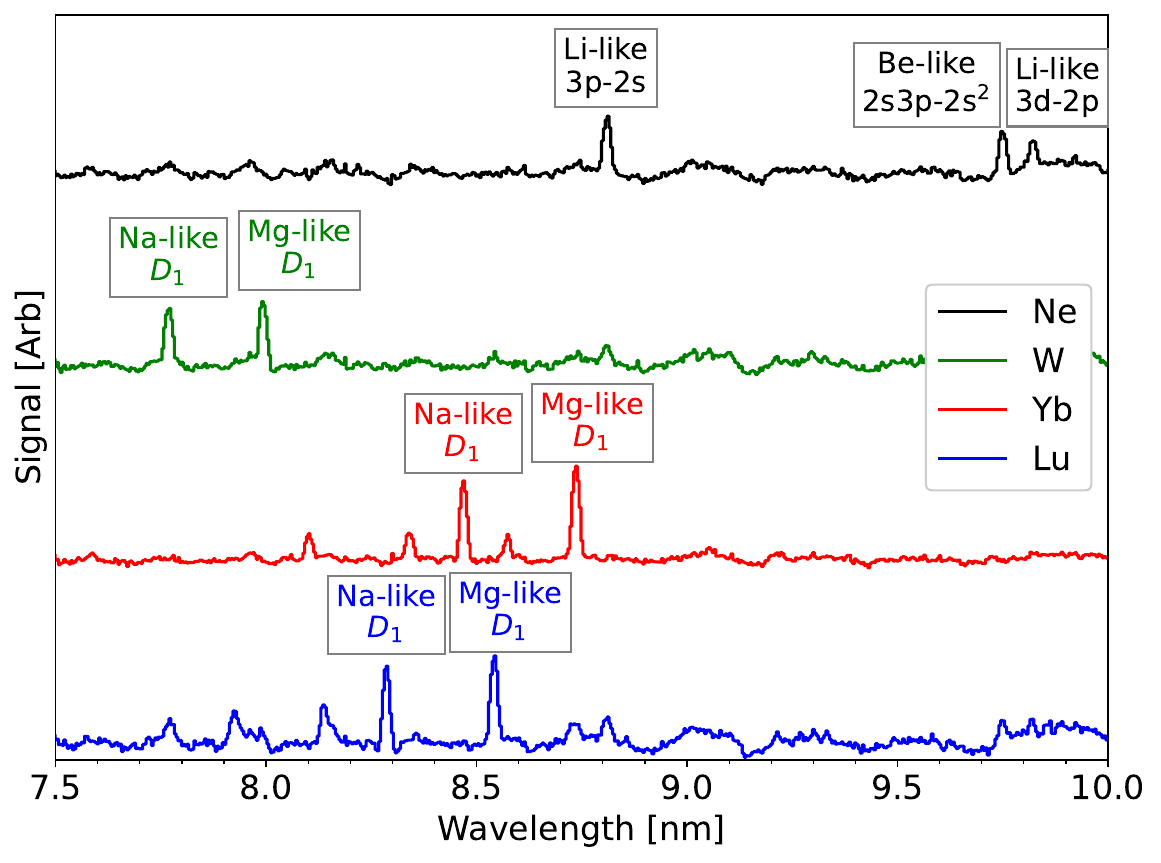}
    \caption{Summed spectra from one day of data collection. The targets of this experiment were the Na-like and Mg-like $D_1$ line separations between Lu and Yb. The $D_1$ lines of W previously measured in \cite{Gillaspy2009} and the Ne lines listed in the NIST ASD \cite{kramida_atomic_2023} were used for calibrating the spectrometer. }
    \label{fig:luyb_spectra}
\end{figure}

\subsection{Measurement Sequence}

The data acquisition campaign was conducted over the course of five days, structured around the operational constraints of the Tokyo EBIT's cryogenic system. The EBIT required liquid helium refills approximately every seven hours, with two fills scheduled per day. This dictated the available measurement windows and segmented the experiment into two approximately 7-hour long sessions per day.

Within each session, spectra were recorded in 10-minute intervals for different injected elements. The sequence was executed in the following order: 30 minutes of ytterbium, 70 minutes of lutetium, 50 minutes of tungsten, and 50 minutes of neon. This sequence maintained roughly equal statistics in the Yb and Lu lines of interest. All of the injected elements were of natural abundance \cite{Berglund2011} due to the challenge of obtaining large enough isotopically pure samples. The Ne spectra were acquired with a 5~Hz ion dump rate to reduce contaminant W species and to provide frequent in situ wavelength calibration references. This alternating sequence ensured that each element was measured under comparable EBIT conditions, including electron beam energy, current, and magnetic field. 

\section{Data Analysis}

The analysis of the extreme ultraviolet spectra focused on extracting the shift in Na-like $3p\,^2P_{1/2} \rightarrow 3s\,^2S_{1/2}$ and Mg-like $3s3p\,^3P_{1} \rightarrow 3s^2\,^1S_{0}$ $D_1$ transition energies between naturally abundant Lu and Yb ions. The data were processed using a combination of image correction, spectral fitting, and wavelength calibration techniques.

\subsection{Spectral Processing and Corrections}

Raw CCD images exhibited several systematic features, including slight tilting of spectral lines and a large background. Line tilt was corrected by rotating the images by $0.8^\circ$. Cosmic ray artifacts were removed using contour detection algorithms, ensuring clean spectra for subsequent analysis~\cite{opencv_library}. To address the time-varying background, we implemented a flat-field correction technique based on a sliding minimum filter. For each pixel, we applied a minimum value filter with a width of 20 pixels, much greater than the line width. The result was then passed through a low pass filter and a Gaussian smoothing filter, and was then subtracted from the original spectra.

\subsection{Line Fitting and Calibration}

Each 10-minute spectrum was analyzed individually. Spectral lines were fit using a Gaussian profile superimposed on a constant background, typically within a narrow window of $\pm 8$ pixels around the peak. Calibration was performed using well-known Ne and W emission lines \cite{kramida_nevii_2006, kramida_neviii_2006, Ralchenko2008, Gillaspy2009}, with lines exhibiting profile complexity or blending excluded from the final calibration fits.

\begin{figure}
    \centering
    \includegraphics[width=0.95\linewidth]{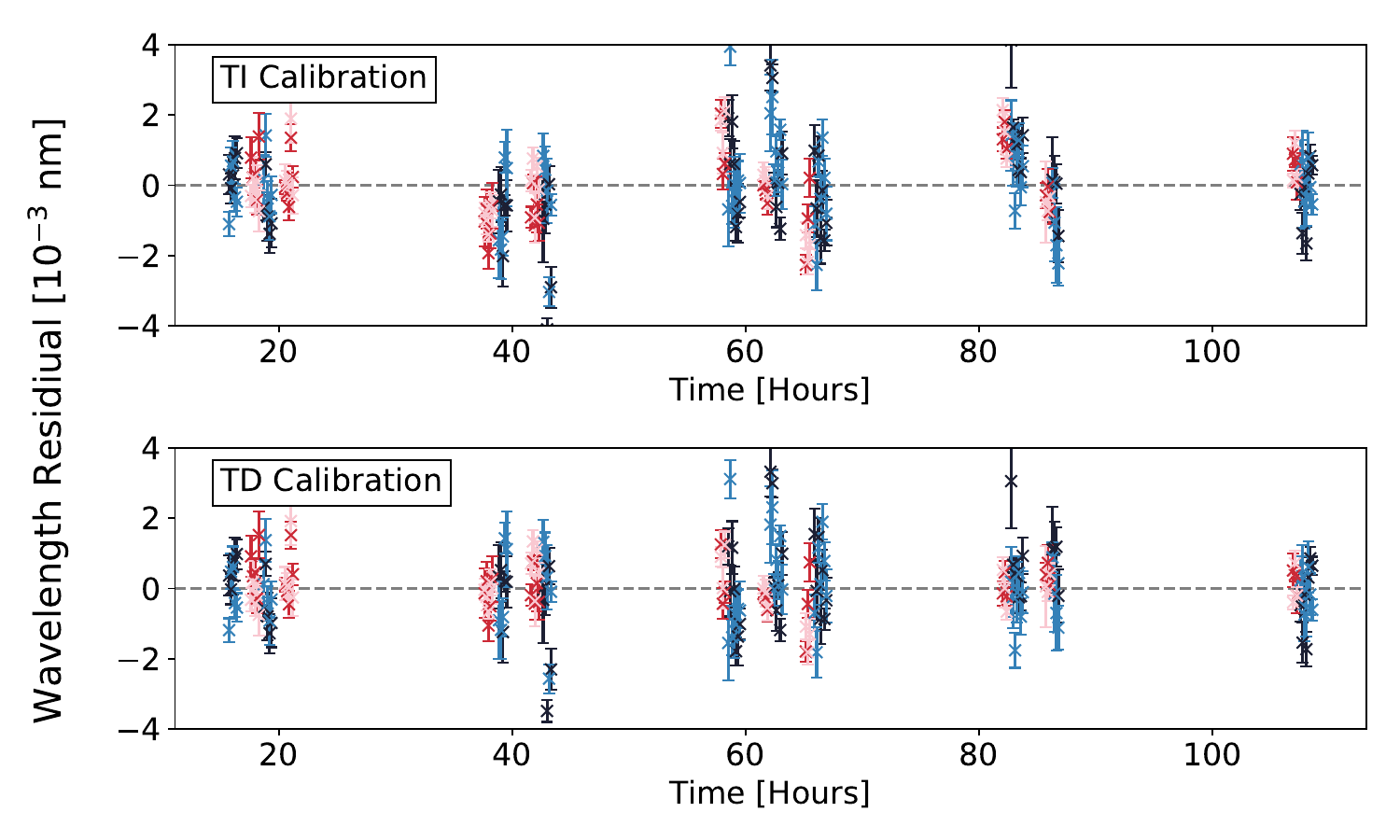}
    \caption{Comparison of the wavelength residuals from a time independent calibration procedure (top) and a time dependent calibration procedure (bottom). The Lu Na-like $D_1$ and Mg-like $D_1$ residuals are shown in dark blue and blue, respectively, with the Yb Na-like and Mg-like residuals shown in red and pink, respectively. Including a linear time dependence over the course of a day removes the majority of the structure in the residuals.}
    \label{fig:cal_dep}
\end{figure}

A time-independent calibration model, based on a quadratic function in pixel space, yielded a reduced $\chi^2$ of 3.2. We attributed this high reduced $\chi^2$ to a visually obvious time dependence of wavelength residuals in Figure \ref{fig:cal_dep}, potentially coming from slight shifts in the electron beam position as the EBIT temperature changed over the course of the day. To improve accuracy, a more refined time-dependent calibration model was implemented which allowed for a linear drift in time over the course of each day, but with a shared spatial dispersion for all days. To assist in determining the time drift, a term in the calibration loss function was added that penalized spread in the calibrated wavelengths of the Na- and Mg-like transitions of Lu and Yb; even though these wavelengths are unknown, they should be consistent over time. The time dependent procedure resulted in a reduced $\chi^2$ of 2.4. 

An F-test \cite{duncan_multiple_1955} comparing the time-independent and time-dependent calibrations showed that the reduction in reduced $\chi^2$ was statistically significant and could not be attributed solely to the additional parameters introduced by the time-dependent model. Increasing the time dependence beyond linear (e.g., quadratic or cubic terms) did not yield a statistically meaningful improvement in fit quality. Therefore, a linear time dependence was adopted.

\subsection{Transition Energy Shift Extraction}

Using the time-dependent calibration and the fitted line positions, the Lu-Yb $D_1$ transition energies shifts were extracted for both Na-like and Mg-like charge states. Both the calibration of the spectrometer and the line statistics contribute to the experimental shift uncertainties. To account for any unresolved systematic uncertainties, all experimental uncertainties were multiplied by the square root of our calibration reduced $\chi^2$ (the ``Birge ratio") \cite{birge_calculation_1932}.

\section{\label{sec:relativistic_theory}Atomic-structure theory}

\subsection{Transition energies of Na- and Mg-like ions}

Transition energies for Na-like ions are calculated using a method \cite{blundell_calculations_1993,Gillaspy2013} that combines third-order RMBPT \cite{johnson_many-body_1988-2} with first-principles QED corrections \cite{blundell_calculations_1993}. As discussed in Ref.~\cite{Gillaspy2013}, various heuristics are employed to estimate omitted QED and many-body contributions, as well as the effects of methodological approximations in some of the terms. Numerical uncertainties arising from grid discretization, basis-set truncation, and integration accuracy are also assessed and are generally found to be negligible. A theoretical uncertainty is then assigned to each transition energy by combining all such effects. This approach yields excellent agreement with experimental $D_1$ transition energies across the whole Na-like isoelectronic sequence \cite{Gillaspy2013,staiger_measurement_2025}, suggesting that the estimated uncertainties---largely unchanged since the early work of Ref.~\cite{blundell_calculations_1993}, when experimental data were sparse---remain reliable indicators of the actual errors. Alternative approaches for the electronic structure of Na-like ions, including multiconfiguration Hartree-Fock calculations \cite{Hosier2025} and an $S$-matrix QED approach \cite{sapirstein_s_2015}, can also be used and produce consistent results for the extracted charge radii \cite{Hosier2025}.

In contrast, the atomic structure of Mg-like ions, which have two valence electrons, is significantly more complex. The $n=3$ valence complex exhibits strong configuration interaction, rendering a straightforward order-by-order RMBPT expansion ineffective. To address this, we adapt a strategy originally developed for Zn-like ions \cite{blundell_calculation_2009}, which face analogous issues in the $n=4$ complex. An effective Hamiltonian for the $n=3$ valence manifold is constructed using third-order RMBPT, supplemented by approximate \emph{ab initio} QED corrections computed in a similar way to those for Na-like ions. The energies of Mg-like states are then obtained by diagonalizing this Hamiltonian. The effective Hamiltonian includes \emph{one-body} terms describing valence-core interactions and \emph{two-body} terms accounting for valence-valence interactions. The one-body terms also occur in the theory of Na-like ions, and the effective Hamiltonian can be conveniently organized so that Na-like transition energies appear along the diagonal. The two-body terms, which are specific to Mg-like ions, contribute to both the diagonal and off-diagonal terms. Zero-body contributions are also present, corresponding to the Ne-like core energy, but these cancel in transition energies and are therefore omitted, reducing the computational complexity substantially. This hybrid CI-RMBPT framework \cite{blundell_calculation_2009} has not yet been applied systematically across the Mg-like isoelectronic series, but very good agreement is found for the Mg-like $D_1$ transition in Yb (see Table~\ref{tab:V}), with an estimated theoretical uncertainty only slightly larger than for the corresponding Na-like $D_1$ transition.

\begin{table*}
\caption{\label{tab:V}Theoretical $D_1$ transition energies of Na- and Mg-like Yb and Lu. The last two columns give the difference of the transition energies (Lu minus Yb). Estimated theoretical uncertainties are shown in parentheses (when different from zero). Notation: ``1-body'', core-valence terms; ``2-body'', valence-valence terms; ``E(0)''--``E(3)'', orders 0--3 of RMBPT including the Coulomb and Breit interactions; ``1body-ret'' and ``2body-ret'', retardation terms; ``Nuc.Rec.'', nuclear-recoil terms; ``SE(val)'', valence self-energy; ``Uehl(val)'', valence Uehling term; ``WK(val)'', valence Wichmann-Kroll term; ``QED(val-x)'', valence-exchange QED (self-energy plus vacuum polarization); ``QED(core)'', core-relaxation QED; ``QED(2-loop)'', phenomenological two-loop QED; ``QED(2-body)'', valence-valence QED in Mg-like ions. Units: eV.}
\begin{ruledtabular}
\begin{tabular}{ldddddd}
\multicolumn{1}{l}{Contribution}  & \multicolumn{2}{c}{Yb\footnotemark[1] ($Z=70$)} &  \multicolumn{2}{c}{Lu\footnotemark[2] ($Z=71$)} & \multicolumn{2}{c}{Lu -- Yb} \\
  & \multicolumn{1}{c}{Na-like $D_1$} & \multicolumn{1}{c}{Mg-like $D_1$} & \multicolumn{1}{c}{Na-like $D_1$} & \multicolumn{1}{c}{Mg-like $D_1$} & \multicolumn{1}{c}{Na-like $D_1$} & \multicolumn{1}{c}{Mg-like $D_1$}\\
\hline
\\
E(0)         &  150.649   &  150.649   &  154.075   &  154.075   &  3.4258    &  3.4258     \\
E(1)         &            &  -4.618    &            &  -4.622    &            &  -0.0043    \\
E(2,1-body)  &  -0.317    &  -0.311    &  -0.321    &  -0.316    &  -0.0042   &  -0.0042    \\
E(2,2-body)  &            &  0.043     &            &  0.039     &            &  -0.0032    \\
E(3)         &  0.004     &  0.004     &  0.004     &  0.004     &  0.0001    &  0.0001     \\
1body-ret    &  0.002(8)  &  0.000(8)  &  0.004(9)  &  0.002(9)  &  0.0021(3) &  0.0021(3)  \\
2body-ret    &            &  0.000     &            &  0.000     &            &  0.0000     \\
Nuc.Rec.     &  -0.007(1) &  -0.007(1) &  -0.007(1) &  -0.007(1) &  0.0000(1) &  0.0000(1)  \\
SE(val)      &  -4.632    &  -4.562    &  -4.895    &  -4.822    &  -0.2624   &  -0.2597    \\
Uehl(val)    &  0.762     &  0.750     &  0.818     &  0.806     &  0.0566    &  0.0559     \\
WK(val)      &  -0.026(2) &  -0.026(2) &  -0.029(2) &  -0.028(2) & -0.0026(2) &  -0.0025(2) \\
QED(val-x)   &  0.057(4)  &  0.056(4)  &  0.060(4)  &  0.059(4)  &  0.0025(2) &  0.0025(2)  \\
QED(core)    &  -0.097(7) &  -0.096(7) &  -0.102(7) &  -0.100(7) & -0.0044(3) &  -0.0044(3) \\
QED(2-loop)  &  0.012(4)  &  0.012(4)  &  0.013(4)  &  0.013(4)  &  0.0009(3) &  0.0009(3)  \\
QED(2-body)  &            &  0.040(12) &            &  0.041(12) &            &  0.0014(4)  \\
\\
Total        & 146.407(13)  & 141.935(17) & 149.621(13) & 145.145(18) &  3.2144(6) &  3.2103(7) \\
Experiment   & 146.399(9)\footnotemark[3]  & 141.938(5)\footnotemark[3] &  &  &  &   \\
\end{tabular}
\footnotetext[1]{Nuclear parameters: $R=5.3051$~fm, $\beta_2=0.320$, $t = 2.3$~fm (see Sec.~\ref{sec:nuclear_radius_determination}).}
\footnotetext[2]{Nuclear parameters: $R=5.3701$~fm, $\beta_2=0.305$, $t = 2.3$~fm (see Sec.~\ref{sec:nuclear_radius_determination}).}
\footnotetext[3]{Reference~\cite{silwal_spectroscopic_2022}}
\end{ruledtabular}
\end{table*}

Theoretical $D_1$ transition energies for Na- and Mg-like Yb and Lu are summarized in Table~\ref{tab:V}. Contributions are separated into RMBPT orders $E(0)$--$E(3)$, where the zeroth-order term $E(0)$ corresponds to the Breit-Coulomb Dirac-Fock (BC-DF) approximation for the Ne-like core, and higher orders of RMBPT also incorporate both Coulomb and Breit interactions. Na-like ions involve only one-body terms, whereas Mg-like ions include both one- and two-body contributions. The QED sector includes valence self-energy, Uehling, and Wichmann-Kroll corrections, as well as approximate treatments of valence-exchange and core-relaxation QED terms. Two-loop QED contributions are incorporated phenomenologically. There are also two-body QED terms specific to Mg-like ions. Further details of the various terms are given in Refs.~\cite{johnson_many-body_1988-2,blundell_calculations_1993,blundell_calculation_2009,Gillaspy2013}.

We note that because the Mg-like transition-energy differences in Table~\ref{tab:V} are obtained by matrix diagonalization, it is not possible to define unique additive contributions such as $E(0)$, $E(1)$, $E(2)$, etc., that sum linearly to yield the final transition energy. Instead, we construct the effective Hamiltonian incrementally: terms are added one at a time, and the matrix is rediagonalized after each addition. The contribution of a given term is then defined as the change in the eigenvalues when that term is included.

It should be emphasized that the contributions obtained this way are not uniquely defined.
Because the diagonalization is nonlinear, the value
attributed to a term may vary slightly depending on the order in which
terms are introduced into the effective Hamiltonian. The first term
added should be $E(0)$, corresponding to differences of BC-DF valence
eigenvalues, since in a perturbative treatment of the diagonalization
these reappear as the familiar energy denominators of RMBPT. The
ordering of the remaining terms is, in principle, arbitrary; here we
adopt the sequence shown in Table~\ref{tab:V} (from top to bottom).
Importantly, the final total, in which all effects are included, is
independent of the order in which the terms are added.

\subsection{Uncertainty in theoretical transition energies}

Most theoretical contributions to the transition energy are electronic in nature and vary smoothly as functions of the nuclear charge $Z$. However, nuclear effects can introduce small irregularities. The dominant such contribution is the nuclear-size correction, which we exploit in this work to constrain nuclear charge radii. Other nuclear effects include nuclear recoil, nuclear polarization, and the hyperfine interaction. 

The recoil correction, which depends on the nuclear mass, is negligible for the present application, amounting to only $0.0000(1)$~eV in the Lu–Yb transition-energy difference (see Table~\ref{tab:V}). We have not explicitly estimated the nuclear-polarization contribution; however, for Li-like ions with $60 \le Z \le 92$, it was shown~\cite{zubova-14} to be comparable in magnitude to the effect of nuclear deformation. As demonstrated in Secs.~\ref{sec:deformation} and \ref{sec:results}, the latter contributes somewhat less than $1$~meV to the transition-energy difference, and we therefore assume nuclear polarization to be negligible for the present purposes.

Regarding the hyperfine effect, if the hyperfine levels in the EBIT are populated statistically, any hyperfine shift is expected to average to zero at the line center. In contrast, a non-statistical population can introduce asymmetries in the line shape and cause a small shift in the measured energy. We estimate this effect to be below $0.1$~meV in the transition-energy difference, rendering it entirely negligible.

\begin{figure}[tb]
\includegraphics[scale=0.78]{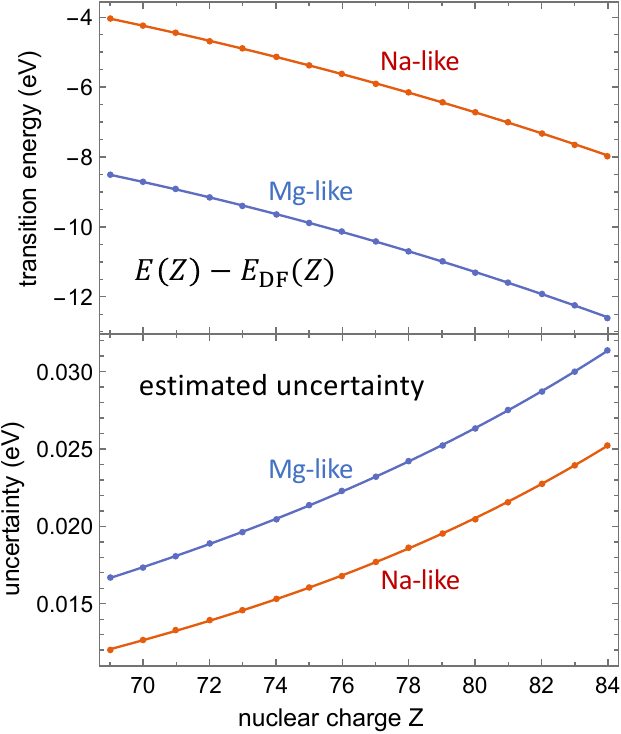}
\caption{\label{fig:smooth_function}Top panel: Difference between the calculated $D_1$ transition energies $E(Z)$ and the corresponding Breit-Coulomb Dirac-Fock values $E_{\text{DF}}(Z)$ for $69\leq Z \leq 84$. Bottom panel: estimated theoretical uncertainties in the transition energies. Dots represent calculated values at each nuclear charge $Z$, and curves show polynomial fits to the data. Na-like $D_1$ transitions are shown in red, Mg-like $D_1$ transitions in blue.}
\end{figure}

In Fig.~\ref{fig:smooth_function}, we show $E_\Delta(Z)=E(Z)-E_{\text{DF}}(Z)$, the difference between the calculated $D_1$ transition energies $E(Z)$ and the corresponding Dirac-Fock baseline value $E_{\text{DF}}(Z)$, which already incorporates the leading nuclear-size correction. Irregularities in $E_\Delta(Z)$ as a function of $Z$ may arise from residual nuclear-size effects, other nuclear contributions, or numerical uncertainties in the calculations. In practice, however, all such deviations are very small, and the data for $E_\Delta(Z)$ in Fig.~\ref{fig:smooth_function} are accurately described by a smooth polynomial fit, denoted $E_{f}(Z)$.

The true error in the calculated transition energy is also dominated by uncalculated electronic terms and is therefore another highly smooth function of $Z$. Our quoted theoretical uncertainties reflect estimates of these omitted higher-order terms. Because purely numerical errors are negligible, the estimated uncertainties vary smoothly with $Z$ and can be parameterized by smooth fitted functions $\Delta E_{f}(Z)$ (see Fig.~\ref{fig:smooth_function}). Note that in this application, we fix the nuclear parameters at ``reference'' values (see Sec.~\ref{sec:nuclear_radius_determination}), so that $E_{\text{DF}}(Z)$ has no uncertainty and $\Delta E_{f}(Z)$ is therefore the estimated uncertainty of the total transition energy $E_f(Z)+E_{\text{DF}}(Z)$.

Now, if the true error in the calculated transition energies $E_\Delta(Z)$ were given by some smooth function $\epsilon(Z)$, the error in the transition-energy difference $E_\Delta(Z)-E_\Delta(Z')$ would be simply $\epsilon(Z)-\epsilon(Z')$. Because $\epsilon(Z)$ is a smooth function, this error is systematic in nature. Noting that the estimated uncertainties $\Delta E_{f}(Z)$ reliably track the true error across the entire isoelectronic sequence, we therefore assume that  $|\epsilon(Z)|$ can be approximated by $\Delta E_{f}(Z)$ and assign an uncertainty of $|\Delta E_{f}(Z)-\Delta E_{f}(Z')|$ to the calculated transition-energy difference. The resulting transition-energy differences between Lu and Yb, along with their estimated uncertainties, are listed in Table~\ref{tab:V}. 

As can be seen from Table~\ref{tab:V}, this procedure generally yields \emph{fractional} uncertainties in the transition-energy differences that are comparable to those of the individual transition energies. However, the \emph{absolute} error in the transition-energy differences is strongly suppressed relative to that in the individual transition energies. Further discussion of these points is given in Appendix~\ref{app:smooth}.

\section{Nuclear Radius Determination\label{sec:nuclear_radius_determination}}

\subsection{\label{sec:scaled_radius}Scaled radius difference}

The extraction of the absolute nuclear charge radius of $^{175}$Lu presented here extends the method recently applied to Ir using Na-like transitions~\cite{Hosier2025}. Here, we exploit the sensitivity of both Na-like ($3p\,^2P_{1/2} \rightarrow 3s\,^2S_{1/2}$) and Mg-like ($3s3p\,^3P_1 \rightarrow 3s^2\,^1S_0$) $D_1$ transitions to nuclear size effects, combining these measurements with high-precision atomic-structure calculations based on relativistic many-body perturbation theory and QED.

We assume a three-parameter Fermi nuclear charge distribution \cite{fricke_nuclear_2004} of the form
\begin{equation}
   \rho(r, \theta, \phi) = \frac{\rho_0}{1+\exp\left[{4\ln(3)(r-c_{\text{def}})}/{t} \right]} 
   \label{eq:nucmod}
\end{equation}
with 
\begin{equation}
    \label{eq:cdef}
    c_{\text{def}}(\theta, \phi) = c [ 1 + \beta_2 Y_{20} (\theta, \phi)]\,.
\end{equation}
The parameter \( t \) is the surface thickness, the distance over which the density falls from 90\% to 10\% of its central value. The parameter \( \beta_2 \) is the quadrupole deformation, with \( Y_{20} \) the relevant spherical harmonic. The constant $\rho_0$ normalizes the distribution such that the total charge equals $Z|e|$. Varying the monopole half-density radius $c$ allows adjustment of the \textit{rms} charge radius, $R = \langle r^2 \rangle^{1/2}$.

In previous works \cite{silwal_determination_2020, Hosier2025}, we handled the effect of higher order nuclear moments on charge radii determinations simply by inflating the uncertainty of the calculated nuclear sensitivity coefficient $S\equiv {\partial{E}}/{\partial{R}}$. We now present an improved formal description that better quantifies the uncertainty in extracted radii due to uncertainties in $t$ and $\beta_2$.

Each choice of nuclear parameters produces a distinct theoretical transition energy, which we denote by $E^T \equiv E^T(R, t, \beta_2)$. For small deviations around a set of reference parameters $(R_0, t_0, \beta_{20})$, the theoretical energy can be approximated by a first-order Taylor expansion,
\begin{align}
    &E^T(R, t, \beta_2) \nonumber \\
    &= E^T(R_0, t_0, \beta_{20}) +
    \delta R \frac{\partial E}{\partial R} 
    + \delta t \frac{\partial E}{\partial t} 
    + \delta \beta_2 \frac{\partial E}{\partial \beta_2} 
   \nonumber \\  
   &\equiv E^T(R_0, t_0, \beta_{20}) + S \delta R + S_t \delta t + S_{\beta_2} \delta \beta_2 \,,\label{eq:higherdiff}
\end{align}
where $S$, $S_t$, and $S_{\beta_2}$ are nuclear \emph{sensitivity} coefficients, defined as the partial derivatives of the transition energy with respect to $R$, $t$, and $\beta_2$, respectively. This linear approximation is valid for sufficiently small parameter variations. Second-order terms such as $\partial^2 E/\partial R^2$, $\partial^2 E/\partial R\partial t$, and $\partial^2 E/\partial R\partial \beta_2$ were evaluated and found negligible in this work (see Sec.~\ref{sec:deformation}).

As in previous works, we focus on transition-energy differences to take advantage of the resulting suppression in both experimental and theoretical uncertainties \cite{silwal_determination_2020, Hosier2025}. Equating theoretical and measured differences for two nuclides $A$ and $B$ yields
\begin{equation}
    \label{eq:matching_diffs}
    E^T_B(R_B, t_B, \beta_{2B}) - E^T_A(R_A, t_A, \beta_{2A})
    = E^M_B-E^M_A \,,
\end{equation}
which, after substituting Eq.~(\ref{eq:higherdiff}), becomes
\begin{align}
D_{BA} \equiv\;& R_{B} - (S_{A}/S_{B}) R_{A} \nonumber\\
= \;& (E^{M}_{B} - E^{M}_{A})/S_{B} \nonumber\\
& - \bigl[E^{T}_{B}(R_{0B}) - E^{T}_{A}(R_{0A})\bigr]/S_{B} \nonumber\\
& + R_{0B} - (S_{A}/S_{B}) R_{0A} \nonumber\\
& - (S_{tB}/S_{B})(t_{B} - t_{0B}) \nonumber\\
& + (S_{tA}/S_{B})(t_{A} - t_{0A}) \nonumber\\
& - (S_{\beta_{2}B}/S_{B})(\beta_{2B} - \beta_{20B}) \nonumber\\
& + (S_{\beta_{2}A}/S_{B})(\beta_{2A} - \beta_{20A})\,,
\label{eq:measurable}
\end{align}
where $E_{B}^{T}(R_{0B})\equiv E_{B}^{T}(R_{0B},t_{0B},\beta_{20B})$. We refer to $D_{BA} \equiv R_{B}-(S_{A}/S_{B})R_{A}$ as the \emph{scaled} radius difference. Equation~(\ref{eq:measurable}) shows that it is the scaled difference, rather than the simple difference $R_B - R_A$, that is directly constrained by a measurement of $E_{B}^{M}-E_{A}^{M}$.

Within the linear approximation of Eq.~(\ref{eq:higherdiff}), an equation of the same form as Eq.~(\ref{eq:measurable}) also holds for two \emph{elements} $A$ and $B$ if all nuclear parameters ($R$, $t$, and $\beta_2$) are replaced by their corresponding isotopic-abundance average. For example, $R_{A}\rightarrow \bar{R}_{A} = \sum_i w_i R_{iA}$, where $w_i$ denotes the isotopic abundance of the $i$th isotope of $A$ with radius $R_{iA}$. This follows from abundance-averaging each side of Eq.~(\ref{eq:measurable}) with respect to $A$ and $B$ for constant reference parameters $(R_{0A}, t_{0A}, \beta_{20A})$ and $(R_{0B}, t_{0B}, \beta_{20B})$. In this case, the reference parameters can be taken to be suitable estimates of the abundance-averaged parameters. Here, we assign $A\rightarrow{}^\text{nat}$Yb and $B\rightarrow{}^{\text{nat}}$Lu, exploiting the accurately known radii and abundances of natural Yb isotopes to determine the natural-abundance-averaged radius of Lu.

Assuming reasonable estimates of the abundance-averaged parameters $\bar{t}$ and $\bar{\beta}_2$ for $A$ and $B$ can be found, we can use these as the reference paramters $t_0$ and $\beta_{20}$, and the final four terms in Eq.~(\ref{eq:measurable}) can then be set to zero. These terms do however contribute to the total uncertainty budget. Propagating uncertainties from Eq.~(\ref{eq:measurable}) yields
\begin{align}
 & \left[\Delta (R_B-\frac{S_A}{S_B} R_A)\right]^2 = \nonumber \\ 
 & \left[\frac{\Delta\left(E^M_B  - E^M_A\right)}{S_B} \right]^2 +
\left\{\frac{\Delta [E^T_B(R_{0B})-E^T_A(R_{0A})]}{S_B} \right\}^2 \nonumber \\
 & {}+ \left\{ \frac{\Delta S_{B}}{S_{B}}[(R_{B}-R_{0B})-(S_{A}/S_{B})(R_{A}-R_{0A})]\right\}^{2} \nonumber \\
 & {}+ \left( \frac{S_{\beta_2B}\Delta \beta_{2B}}{S_B} \right)^2 
+ \left( \frac{S_{tB} \Delta t_B}{S_B} \right)^2 \nonumber \\
 & {}+ \left( \frac{S_{\beta_2A} \Delta \beta_{2A}}{S_B} \right)^2 
+ \left( \frac{S_{tA} \Delta t_A}{S_B} \right)^2\,.
\label{eq:radius_uncert}
\end{align}
The last four terms arise from the uncertainties in  $\bar{t}$ and $\bar{\beta}_2$. The third term arises from the uncertainty in the sensitivity coefficient $S_B$ (due to omitted terms and numerical error in the atomic theory), while the analogous terms arising from the uncertainties in $S_t$ and $S_{\beta_2}$ are negligible and have been omitted. Note that $\Delta [E^T_B(R_{0B})-E^T_A(R_{0A})]$ represents the uncertainty in the theoretical energy difference assuming fixed nuclear parameters---the quantity listed in Table~\ref{tab:V} and discussed in Sec.~\ref{sec:relativistic_theory}.

This framework enables systematic uncertainty propagation and provides a robust methodology for extracting nuclear charge radii from high-precision spectroscopic measurements.

\subsection{Sensitivity to Nuclear Parameters\label{sec:deformation}}

Radii extracted using the HCI technique are largely insensitive to variations in $t$ and $\beta_2$ when the charge radius $R$ is fixed. To illustrate this, we determined sensitivity coefficients by evaluating the theoretical transition energies for five values of each nuclear parameter centered on its reference value, followed by polynomial fitting. For the radius sensitivity coefficient $S$, the energies were calculated using the full theoretical framework (RMBPT plus QED). The other two coefficients, $S_t$ and $S_{\beta_2}$, primarily affect only the uncertainty through Eq.~(\ref{eq:radius_uncert}); for these, the simpler BC-DF level of theory was employed. The resulting sensitivity coefficients are summarized in Table~\ref{tab:sens_coef}.

\begin{table}[htbp!]
    \centering
    \caption{Sensitivity coefficients for the Na-like and Mg-like $D_1$ transitions. The radius sensitivity $S$ was calculated at the full RMBPT+QED level, while the skin-thickness sensitivity $S_t$ and the deformation sensitivity $S_{\beta_2}$ were evaluated at the BC-DF level.}
    \begin{tabular}{c@{\hskip 0.15 in}c@{\hskip 0.05 in}c@{\hskip 0.15 in}c@{\hskip 0.05 in}c}
        \toprule
         & \multicolumn{2}{@{\hskip 0.42 in}l}{Yb} & \multicolumn{2}{c}{Lu} \\
         \rule{0pt}{0.15in} & Na-like & Mg-like & Na-like & Mg-like  \\
         \midrule
        $S$ [eV/fm] & $-0.224(1)$ & $-0.218(1)$ & $-0.250(1)$ & $-0.243(1)$ \\ 
        $S_t$ [eV/fm] & 0.0010 & 0.0010 & 0.0012 & 0.0012 \\ 
        $S_{\beta_2}$ [eV] & 0.0024 & 0.0024 & 0.0027 & 0.0027 \\ 
        \bottomrule
    \end{tabular}
    \label{tab:sens_coef}
\end{table}

As shown in Sec.~\ref{sec:results}, the contributions of $t$ and $\beta_2$ to the total uncertainty are very small. This outcome is already suggested by the relative magnitudes of the sensitivities to $R$, $t$, and $\beta_2$ in Table~\ref{tab:sens_coef}, indicating that even under weak constraints on the latter parameters, the radius difference can still be determined with high precision.

We also consider the second-order size correction to Eq.~(\ref{eq:higherdiff}), $\delta E = S\delta R + S_2 (\delta R)^2$ (for $\delta t = \delta \beta_2 = 0$). From the polynomial fit at BC--DF level, we obtain second-order sensitivity coefficients $S_2 = -0.015$~eVfm$^{-2}$ for Yb and $S_2 = -0.016$~eVfm$^{-2}$ for Lu. Our final inferred shift in the Lu radius is $\delta R \approx 0.08$~fm relative to the reference value (see Sec.~\ref{sec:results}), implying a second-order energy correction of $S_2 (\delta R)^2 \approx 0.1$~meV, which is negligible. Likewise, the radius difference between $^{168}$Yb and $^{176}$Yb is approximately $0.05$~fm~\cite{Angeli13}, indicating that the abundance averaging of Eq.~(\ref{eq:higherdiff}) also introduces negligible error.

To investigate the effect of higher-order terms such as  $\partial^2 E/\partial R\partial t$ and $\partial^2 E/\partial R\partial \beta_2$ in Eq.~(\ref{eq:higherdiff}), we evaluated $S$ at different values of $\beta_{20}$ and $t_0$. The results in Table \ref{tab:highersens} reveal that $S$ fluctuates near the 0.01\% level, a negligible amount relative to the 0.5\% assumed uncertainty in $S$ due to omitted terms in the atomic-structure theory.

\begin{table}[htbp!]
    \centering
    \caption{Variations in $S^\text{Na}_\text{Yb}$ calculated at the BC-DF level for different nuclear model parameters. Unless otherwise indicated, calculations assume $R = 5.3051\,\text{fm}$, $\beta_{2}=0.32$, $t=2.3\,\text{fm}$.}
    \begin{tabular}{c@{\hskip 0.1 in}c}
        \toprule
        & $S$ [eV/fm] \\
        \midrule
        & $-0.224367$\\
        $t=2.1\,\text{fm}$ & $-0.224350$\\
        $t=2.3\,\text{fm}$ & $-0.224387$\\
        $\beta_2 = 0.30$ & $-0.224386$\\
                $\beta_2 = 0.34$ & $-0.224346$\\
        \bottomrule
    \end{tabular}
    \label{tab:highersens}
\end{table}

For completeness, the adopted abundance-averaged values of $t$ and $\beta_2$ for all naturally abundant Yb and Lu isotopes are provided in Table~\ref{tab:def_pars}.

\begin{table}[htbp!]
    \centering
    \caption{Adopted skin thicknesses, deformation parameters, and natural abundances for the isotopes considered in this work. Skin thicknesses are conservative estimates based on data from other rare-earth elements \cite{fricke_nuclear_1995}. The natural-abundance deformation parameters are computed as abundance-weighted averages across all isotopes, assuming complete error correlation within each isotopic chain.}
    \begin{tabular}{l@{\hskip 0.1 in}c@{\hskip 0.1 in}l@{\hskip 0.1 in}c}
        \toprule
          & $t$ [fm] & \multicolumn{1}{c}{$\beta_2$ [-]} & Abundance\footnote{Reference \cite{Berglund2011}} \\
         \midrule
         $^{168}$Yb & & 0.322(9)\footnote{Reference \cite{Raman2001, Pritychenko2016}} & 0.126\% \\
         $^{170}$Yb & & 0.326(4)\textsuperscript{b} & 3.02\% \\
         $^{171}$Yb & & 0.328(5)\footnote{Interpolated from neighboring isotopes} & 14.2\% \\
         $^{172}$Yb & & 0.330(2)\textsuperscript{b} & 21.8\% \\
         $^{173}$Yb & & 0.328(4)\textsuperscript{c} & 16.1\% \\
         $^{174}$Yb & & 0.325(2)\textsuperscript{b} & 31.9\% \\
         $^{176}$Yb & & 0.305(5)\textsuperscript{b} & 12.6\% \\
         $^{175}$Lu & & 0.31(3)\footnote{Reference \cite{GBK98}} & 97.4\% \\
         $^{176}$Lu & & 0.30(3)\textsuperscript{d} & 2.6\% \\
         \midrule
         $^{\text{nat}}$Yb & 2.3(2) & 0.320(3) & \\
         $^{\text{nat}}$Lu & 2.3(2) & 0.305(30) & \\
         \midrule
    \end{tabular}
    \label{tab:def_pars}
\end{table}

We note that the present uncertainty estimates do not account for any dependence on the assumed nuclear model beyond the three-parameter Fermi distribution, Eq.~(\ref{eq:nucmod}). It is improbable that the actual charge distribution is perfectly described by a Fermi function, as model-independent electron-scattering data demonstrate \cite{devries_nuclear_1987}. Nevertheless, Table~\ref{tab:sens_coef} makes clear that the dominant effect on the $D_1$ transition energies is the charge radius, rather than higher-order nuclear moments. While a fully model-independent determination of radius differences from HCI transitions will be pursued in future work, we consider it justified at the current level of precision to neglect nuclear-model dependence.

\section{Nuclear Structure Theory\label{sec:nuc_theory}}

We compare our results with predictions from nuclear DFT. For this work, we have selected two relevant energy density functionals. As a baseline, we employ the conventional Skyrme functional \cite{Bender2003} with the SV-min parametrization \cite{Kluepfel2009}. We also use the Fayans functional \cite{Fayans1998, Fayans2000}, which includes gradient surface and pairing terms. Specifically, we use the two parametrizations Fy($\Delta$r, HFB)~\cite{Miller2019} and  Fy(IVP) \cite{Karthein2024}.

While effective for global properties, standard Skyrme functionals like SV-min struggle to reproduce subtle isotopic trends, such as the specific curvature observed in charge radii between shell closures~\cite{Reinhard2017global}. The inclusion of additional gradient terms in the surface energy and pairing fields of the Fayans functional has been seen to provide the flexibility needed to describe isotopic shifts and isotopic OESR. The Fy($\Delta$r, HFB) enforces identical pairing strengths for both protons and neutrons. Fy(IVP) extends this by allowing for different pairing strengths for protons and neutrons. This separation allows for a specific tuning of neutron pairing and helps in modeling the competition between nuclear deformation and the spherical-driving force of pairing, particularly in regions like neutron-deficient indium isotopes~\cite{Karthein2024}.

Calibration of these DFT models were done with the dataset from Ref. \cite{Kluepfel2009}. Additionally the Fy($\Delta$r, HFB) calibration dataset includes measurements along the Ca isotopic chain and the Fy(IVP) calibration dataset also contains measurements along the semimagic Sn (Z = 50) and Pb (Z = 82) isotopic chains. The deformed nuclei are computed with axial DFT solver \texttt{SkyAx} \cite{SkyAx2021}. To estimate model uncertainties we propagate parameter covariances through the linear regression formalism so that both central predictions and uncertainty bands are provided for radii and for OES diagnostics~\cite{Kluepfel2009,Dob14}.

Odd-$A$ nuclei were computed using the self-consistent blocking method in the Hartree-Fock-Bogoliubov (HFB) framework~\cite{(Rin80b)}: candidate one-quasiparticle configurations are identified by their parity and K quantum numbers and individually blocked to obtain their converged energy and density. Experimental one-quasiparticle states~\cite{Nazarewicz1990} were used for this. Charge densities were constructed from the proton mean-field densities including standard finite-size corrections, Darwin and spin–orbit relativistic terms, and nucleon form-factor contributions~\cite{DFTformfactors}. We compute root-mean-square charge radii from these densities so that our theoretical radii are directly comparable to the experimental radii.

\begin{figure}[t]
      \includegraphics[width=0.95\columnwidth]{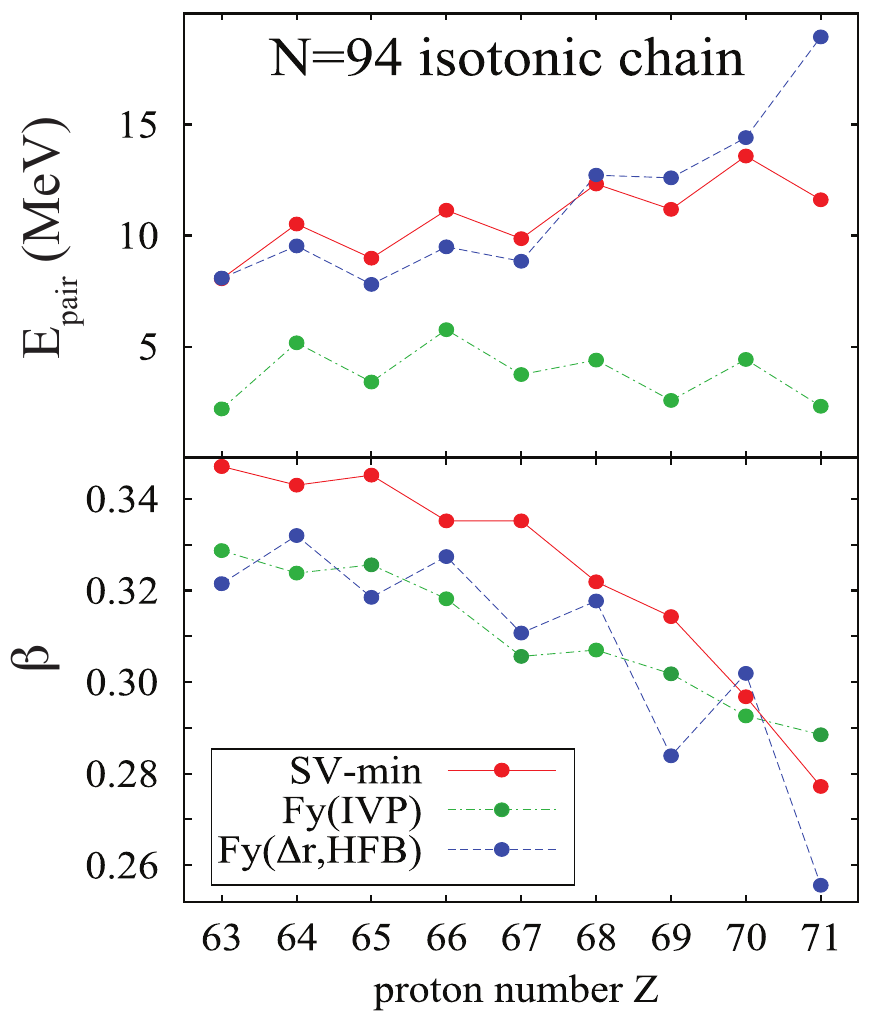}
\caption{DFT predictions for the $N = 94$ isotonic chain for three EDFs parameterizations as indicated.
Top: Pairing energy; Bottom: Quadrupole deformation $\beta$.}
\label{fig:Predictions}
\end{figure}

Figure \ref{fig:Predictions} shows predicted OES of pairing energy (expectation value of the paring energy density functional) and quadrupole deformation for the $N=94$ isotones. The pairing energy $E_{\rm pair}$ is not observable, and it shows significant model dependence while all the EDF parametrizations used reproduce very well the observed OES of binding energy. When moving along the isotonic $N=94$ chain, the staggering in $E_{\rm pair}$ is weak; it reflects the usual reduction of pairing correlations in odd-$A$ nuclei due to blocking. Since pairing   is a symmetry-restoring correlation \cite{Reinhard1984,Nazarewicz1994}, larger values of  $E_{\rm pair}$ are expected to reduce nuclear deformations, and vice-versa.

The quadrupole deformation $\beta$ is deduced from the calculated proton quadrupole moment, and can be compared to experimental quadrupole moments, see Ref.~\cite{Reinhard2022} for examples. The OES of  quadrupole deformations is also weak: small fluctuations of $\beta$ seen in odd-$Z$ reflect different shape-polarization effects of different one-quasiproton orbits 
\cite{Nazarewicz1990,Reinhard2022} and reduced pairing. 

\section{Results\label{sec:results}}

\begin{figure}
    \centering
    \includegraphics[width=0.99\linewidth]{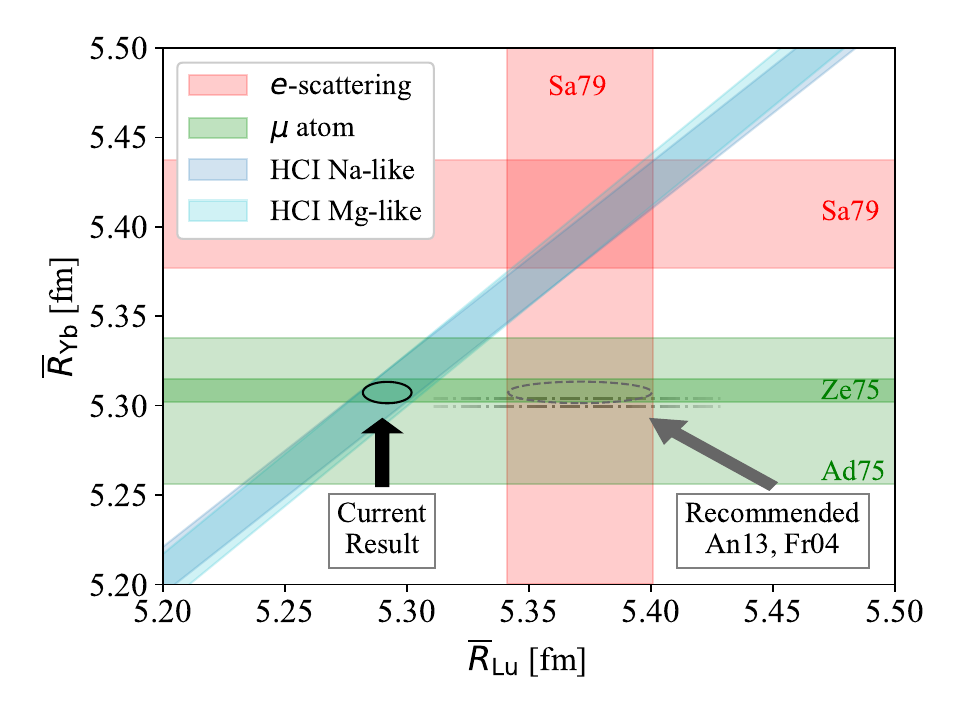}
    \caption{1-$\sigma$ constraints placed on the joint probability distribution of naturally abundant Yb and Lu. Previous absolute measurements from electron scattering (Sa79: \cite{Sasanuma1979}) and muonic atom spectroscopy (Ze75: \cite{zehnder_charge_1975}, Ad75: \cite{adler_study_1975}) are displayed in red and green bands, respectively. The 1-$\sigma$ confidence region An13 \cite{Angeli13} is shown by the gray ellipse, with the 1-$\sigma$ confidence region from Fr04  \cite{fricke_nuclear_2004} (which does not list an uncertainty for the Lu radii) shown by the horizontal gray lines. The HCI constraints from this work are shown in blue for Na-like and cyan for Mg-like, with our current recommendation for the Lu and Yb radii represented by a black ellipse.}
    \label{fig:lu_yb}
\end{figure}

We determined the scaled difference in radius between natural-abundance Yb and Lu using extreme ultraviolet spectroscopy of Na-like and Mg-like ions. As input to the determination of the radius difference, we measured the following energy shifts:
\begin{align*}
&E^{M,\text{Na}}_\text{Lu} - E^{M,\text{Na}}_\text{Yb} = 3.235(3)\,\text{eV} \\
&E^{M,\text{Mg}}_\text{Lu} - E^{M,\text{Mg}}_\text{Yb} = 3.229(3)\,\text{eV}.
\end{align*}
The theoretical transition energies were calculated  using arbitrary reference radii of $\bar{R}_{\text{Yb}_0} = 5.3051$\,fm and $\bar{R}_{\text{Lu}_0} = 5.3701$\,fm. These were determined by taking natural-abundance averages of the values from the recommended nuclear charge radii of Ref.~\cite{Angeli13}. Note that within the linear approximation of Eq.~(\ref{eq:higherdiff}), the precise values assumed for $\bar{R}_{\text{Yb} 0}$ and $\bar{R}_{\text{Lu} 0}$ do not affect the results. Abundance-averaged skin thicknesses and deformation parameters were taken from Table \ref{tab:def_pars}. The theoretical natural abundance averaged energies of Lu and Yb are (see Table~\ref{tab:V}):
\begin{align*}
E^{T\text{,Na}}_\text{Lu}(\bar{R}_{\text{Lu}_0}) - E^{T\text{,Na}}_\text{Yb}(\bar{R}_{\text{Yb}_0}) &= \text{{3.2144(6)}}\,\text{eV}, \\
E^{T\text{,Mg}}_\text{Lu}(\bar{R}_{\text{Lu}_0}) - E^{T\text{,Mg}}_\text{Yb}(\bar{R}_{\text{Yb}_0}) &= \text{{3.2103(7)}}\,\text{eV} .
\end{align*}

\begin{table}[]
    \caption{Error budget for the Na-like and Mg-like determinations of the Yb-Lu scaled radius differences [Eqs.~(\ref{eq:nalike_r}) and (\ref{eq:mglike_r})]. The center column indicates the correlation coefficient between the two determinations for that term.}
    \centering
\begin{tabular}{c@{\hskip 0.1 in}c@{\hskip 0.25 in}c@{\hskip 0.25 in}c}
    \toprule
    \begin{tabular}{@{}c@{}}Variance \\ Contribution\end{tabular} &
    \begin{tabular}{@{}c@{}}Na-like\\ $[10^{-6}\,\text{fm}^2]$\end{tabular} &
    $\rho_{\text{Na}, \text{Mg}}$ &
    \begin{tabular}{@{}c@{}}Mg-like\\ $[10^{-6}\,\text{fm}^2]$\end{tabular} \\
        \midrule
        Lu, Stat & 64.3 & 0.00 & 69.0\\
        Yb, Stat  & 56.7 & 0.00 & 61.2\\
        Calibration  & 15.2 & 0.97 & 14.5\\
        \textbf{Experiment Total} & 136.3 & 0.10 & 144.7 \\
        \rule{0pt}{0.2in}$E^T_B(R_{B_0})-E^T_A(R_{A_0})$ & 5.8 & 0.70 & 8.3 \\
        $S$ & 0.1 & 1.00 & 0.1\\
        \textbf{Theory Total} & 5.9 & 0.70 & 8.4 \\
        \rule{0pt}{0.2in}$t_\text{Lu}$ & 0.9 & 1.00 & 1.1\\
        $\beta_{2, \text{Lu}}$ & 0.1 & 1.00 & 0.1\\
        $t_\text{Yb}$ & 0.6 & 1.00 & 0.7\\
        $\beta_{2, \text{Yb}}$ & 0.0 & 1.00 & 0.0\\
        \textbf{Parameters Total} & 1.6 & 1.00 & 1.9\\
        \rule{0pt}{0.2in}\textbf{Total} & 143.8 & 0.15 & 155.0 \\
        \bottomrule
    \end{tabular}
    \label{tab:error_budget}
\end{table}

Inserting these into Eq.~(\ref{eq:measurable}) yields the following constraints on the Yb and Lu nuclear radii
\begin{align}
      \bar{D}^{\text{Na}}_{\text{Lu,Yb}}=\bar{R}_\text{Lu} - \frac{S^{\text{Na}}_\text{Yb}}{{S^\text{Na}_\text{Lu}}}\bar{R}_\text{Yb} &= 0.535(12)~\text{fm} 
      \label{eq:nalike_r}
    \\
    \bar{D}^{\text{Mg}}_{\text{Lu,Yb}}=\bar{R}_\text{Lu} - \frac{S^{\text{Mg}}_\text{Yb}}{S^\text{Mg}_\text{Lu}}  \bar{R}_\text{Yb} &= 0.540(13)~\text{fm}.
    \label{eq:mglike_r}
\end{align}
According to our calculations, the ratios $S^\text{Na}_\text{Yb}/S^\text{Na}_\text{Lu}$ and $S^\text{Mg}_\text{Yb}/S^\text{Mg}_\text{Lu}$ differ from each other by less than one part in 10$^4$ (with $S^\text{Na}_\text{Yb}/S^\text{Na}_\text{Lu}=0.896$). Therefore, Eqs.~(\ref{eq:nalike_r}) and (\ref{eq:mglike_r}) imply that the Na- and Mg-like measurements yield mutually consistent Lu-Yb radius differences.  The constraints in Eqs.~(\ref{eq:nalike_r}) and (\ref{eq:mglike_r}) are illustrated in Fig.~\ref{fig:lu_yb}, along with all of the previously discussed muonic atom spectroscopy and elastic electron scattering results. Given the natural-abundance-averaged nature of our work, we converted previous measurements to their natural-abundance-averaged equivalent using the abundances of Ref.~\onlinecite{Berglund2011} and the optical isotope shift (OIS) studies of Ref.~\onlinecite{GBK98} and Refs.~\onlinecite{Barzakh1998, Barzakh2000, Schulz1991, Jin1991, Sprouse1989}. 

The uncertainty in $\bar{D}^{\text{Na}}_{\text{Lu,Yb}}$ and $\bar{D}^{\text{Mg}}_{\text{Lu,Yb}}$ is dominated by statistical uncertainties and residual calibration uncertainties, as shown by Table \ref{tab:error_budget}. The next largest source of uncertainty is the theoretical energy difference for fixed nuclear parameters (4\%--5\% of total variance), followed by the uncertainties associated with the skin thicknesses and deformation parameters (1\% of total variance).

To interpret our result in terms of the radii of specific isotopes, a generalized least squares fit is necessary that considers the strong correlations present in the data. This was accomplished using a preliminary version of the \texttt{radbase} set of codes \cite{iaea_summary_2025}, which treats the isotopic radii as parameters and optimizes them given a set of measurements. In addition to the constraints of Eqs.~(\ref{eq:nalike_r}) and (\ref{eq:mglike_r}), we included the OIS radius differences of Ref.~\onlinecite{GBK98} for Lu, and those of Ref.~\onlinecite{kawasaki_isotopeshift_2024} for Yb. For both OIS works, a high degree of correlation between the different radius differences was assumed, as the main source of uncertainty was the theoretical uncertainty in the field shift coefficient. The primary challenge of this analysis is selecting the measurement(s) to use for the absolute radius of individual Yb and Lu isotopes. We completed the optimization for various choices of included absolute measurements, and present the results in Table \ref{tab:opt_results}. For a description of the optimization procedure, see Appendix \ref{app:optimization}.

We note that, although there is a breadth of OIS data available for Lu and Yb, the OIS results of \cite{GBK98} and \cite{kawasaki_isotopeshift_2024} are sufficient for the purpose of determining the isotonic differences and absolute radii. The absolute uncertainties of OIS work are small enough that neglecting them entirely does not affect any of the results in this work by more than 1\% of the reported uncertainty.

\begin{table}
    \caption{The $^{175}$Lu radius and $^{175}$Lu-$^{174}$Yb isotonic radius difference determined from this work using various sources of the Yb radii. Each row represents the combination of this work (Eqs. \ref{eq:nalike_r} and \ref{eq:mglike_r}), the optical isotope shift measurements of \cite{GBK98} and \cite{kawasaki_isotopeshift_2024}, and different Yb radii. The optimization procedure to obtain the presented values is described in detail in Appendix \ref{app:optimization}. An13: \cite{Angeli13}, Fr04: \cite{fricke_nuclear_2004}, Ze75: \cite{zehnder_charge_1975}, Sa79 \cite{Sasanuma1979}.}
    \begin{tabular}{rl@{\hskip 0.1 in}c@{\hskip 0.1 in}c}
        \toprule
        \multicolumn{2}{c}{Reference Radius~[fm]} & $R_{^{175}\text{Lu}}$~[fm]&$R_{^{175}\text{Lu}}-R_{^{174}\text{Yb}}$~[fm] \\
        \midrule

        An13: $\bar{R}_{\text{Yb}}$ = &5.306(6)  & 5.291(11) & $-0.017(10)$ \\ 
        Fr04: $\bar{R}_{\text{Yb}}$ = &5.302(2)  & 5.288(10) & $-0.017(10)$\\
        Ze75: $R_{^{174}\text{Yb}}$ = &5.308(6) & 5.290(11) & $-0.017(10)$\\
        Sa79: $R_{^{174}\text{Yb}}$ = &5.410(30) & 5.382(30) & $-0.028(10)$\\
        \bottomrule
    \end{tabular}
    \label{tab:opt_results}
\end{table}

\begin{figure}[t]
   \centering
   \includegraphics[width=0.95\linewidth]{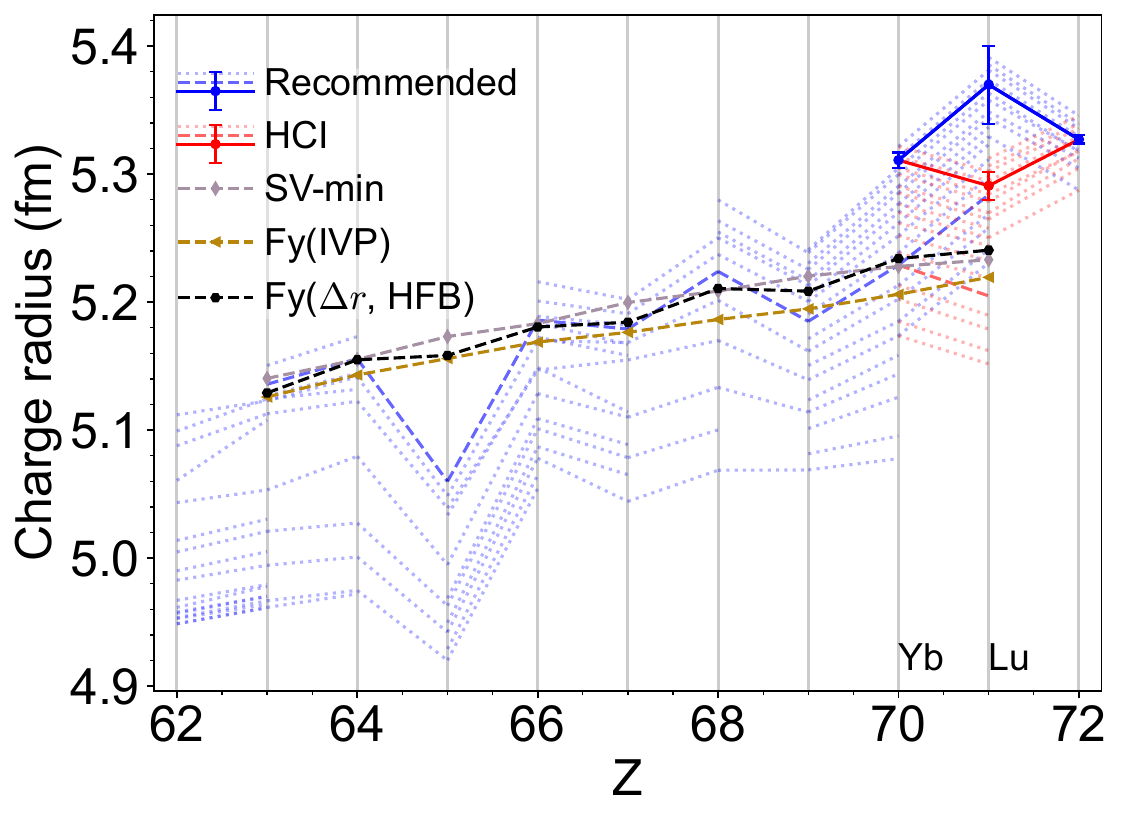}
 \caption{Restored odd-even staggering with the measured nuclear charge radius difference between ytterbium and lutetium. Red symbols show results obtained using the new anchor reference radius $R_{^{175}\text{Lu}} = 5.291(11)~\text{fm}$ for the OIS difference measurements \cite{GBK98}, while blue symbols and lines represent recommended values from Angeli and Marinova \cite{Angeli13}. The data points with solid lines and error bars correspond to the \(N=104\) isotonic sequence, which includes the \textsuperscript{175}Lu isotope. Density functional theory predictions using functionals described in Section \ref{sec:nuc_theory} and radii from Angeli and Marinova for the \(N=94\) isotonic chain are shown in dashed lines.}

   \label{fig:odd_even_abs}
 \end{figure}
 
\section{Discussion}

\begin{figure}[t]

      \includegraphics[width=0.9\columnwidth]{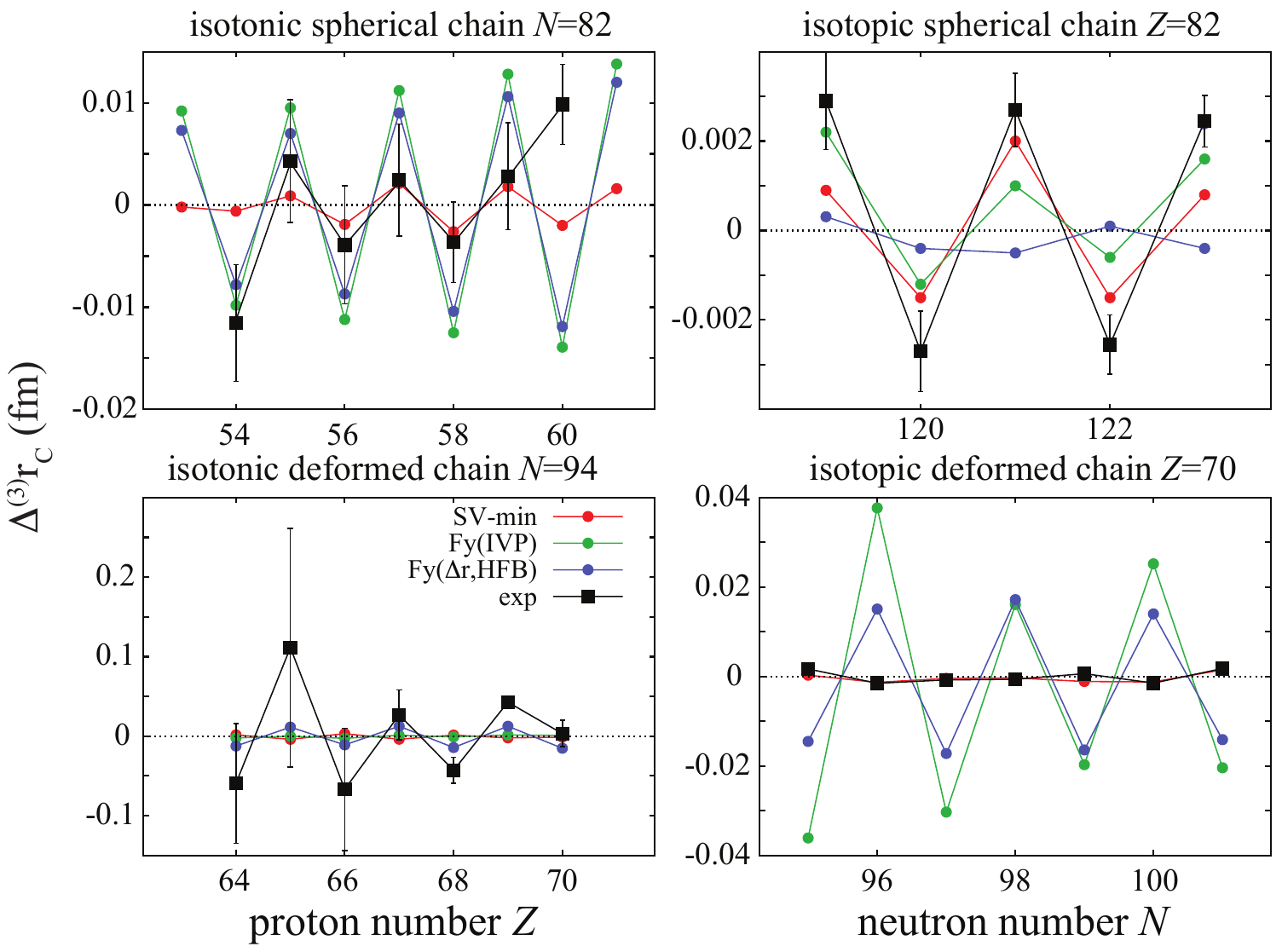}
\caption{DFT predictions and experimental data \cite{Angeli13} for $\Delta_R^{(3)}$ in isotonic (left) and isotopic (right) chains of selected spherical (top) and deformed (bottom) nuclei.}
\label{fig:OES-SD}
\end{figure}

The extraction of the Lu-Yb scaled radius difference using EUV spectroscopy of Na-like and Mg-like ions demonstrates the precision and robustness of our method. By combining high-resolution spectroscopic measurements with relativistic many-body perturbation theory and a generalized least-squares optimization framework that includes muonic atom and OIS data, (Appendix~\ref{app:optimization}), we achieve a threefold reduction in uncertainty compared to previous electron scattering results~\cite{Sasanuma1979}.

The agreement between the Na-like and Mg-like determinations validates the reliability of our calibration model and theoretical sensitivity coefficients, and places a unique constraint on the set of Lu and Yb charge radii. The time-dependent calibration model introduced here significantly improves wavelength stability and reduces systematic drift, as shown in Fig.~\ref{fig:cal_dep}.

Table \ref{tab:opt_results} reveals that applying the constraints of Eqs.~(\ref{eq:nalike_r}) and (\ref{eq:mglike_r}) results in normal odd--even staggering between Lu and Yb, regardless of the absolute Yb measurements used. This consistency in the $^{175}$Lu--$^{174}$Yb radius difference reinforces that our HCI based method allows for near-direct studies of isotonic trends, even when the absolute radius data has large uncertainties or disagreements. 

Using our method, the absolute radius of $^{175}$Lu (and the other Lu isotopes) depends on the value of the absolute nuclear charge radius of $^{174}$Yb used as the anchor. The discrepancy between the values in Table \ref{tab:opt_results} demonstrate that there is conflict between the electron scattering and muonic atom spectroscopy data in Fig.~\ref{fig:lu_yb} that is not resolved by our measurement. However, given the agreement of muonic atom results, and the effective adoption of the radius from muonic atom spectroscopy in previous compilations \cite{fricke_nuclear_2004, Angeli13}, we use first row of Table \ref{tab:opt_results} to make the assignment of $R_{^{175}\text{Lu}} = 5.291(11)~\text{fm}$. The tension between muonic atom and electron scattering experiments warrants a reevaluation of the underlying data, potentially using theoretical techniques discussed in \cite{iaea_summary_2025}. Our change to the \textsuperscript{175}Lu absolute nuclear charge radius also implies a shift in the rest of the Lu isotopic chain, which we show alongside the DFT calculations of the $N=94$ isotonic sequence in Fig. \ref{fig:odd_even_abs}.

 We discuss the implications of this restoration of OESR in a companion paper~\cite{companionTakacs2025}, but note here that the observed reduction in the radius of $^{175}\mathrm{Lu}$ is consistent with proton pairing effects and reduced surface diffuseness found in odd $Z$ nuclei. Our nuclear DFT calculations, however, predict a much smaller OESR than both the currently recommended values along the N=94 isotonic sequence \cite{Angeli13} and our experimental result. To illustrate this discrepancy, Fig.~\ref{fig:OES-SD} compares experimental and theoretical values of $\Delta_R^{(3)}$ for the isotonic chains at $N=82$ and $92$ and for the isotopic chains at $Z=82$ and $70$. The magnitude of the OESR for the spherical sequences and for the deformed Yb isotopic chain are consistent with DFT predictions, but the experimental OESR for the $N=94$ nuclei is exceptionally large when compared to DFT calculations.

The framework presented here is broadly applicable to future studies of isotonic and isoelectronic sequences. The Mg-like calculation presented here allows for a cross check of our result and improves the statistical uncertainty achievable. Our new approach to nuclear model dependence (explicitly accounting for the uncertainty in $t$ and $\beta_2$ rather than inflating the uncertainty of $S$ as done in \cite{Hosier2025}), results in a much more rigorous treatment of the nuclear shape. 

\section*{Conclusion}

We presented a refined determination of the Yb-Lu natural abundance averaged scaled radius difference. Combining our work with results from muonic atom spectroscopy and optical isotope shift studies, we determined that the $^{175}$Lu radius was smaller than the $^{174}$Yb radius, in direct contrast to the radii in previous nuclear radius compilations. Our measurement provides the first constraint on the Lu radii since the 1979 electron scattering result of Sasanuma, reducing the experimental uncertainty of the $^{175}$Lu radius by a factor of three. The new value of $R_{^{175}\text{Lu}} = 5.291(11)~\text{fm}$ restores the expected odd--even staggering along isotonic chains involving the Lu isotopes.

The Mg-like calculation and improved nuclear shape handling presented here will allow for a reanalysis of previous HCI work \cite{Hosier2025} and yield reduced uncertainties in future experiments. Further reductions in uncertainty will have to come from studies using isotopically enriched samples, increased statistics, new EUV and x-ray measurement techniques, or lines from different charge states.

Future experiments will focus on the rest of the nuclides in the deformed region as well as on high-$Z$, radioactive nuclides. For elements with $Z>50$, HCI experiments are a prime candidate to directly study isotonic nuclear charge radius systematics and supplement muonic atom and electron scattering data. Given the tension between electron scattering and muonic atom results, includingcross-element constraints, HCI measurements provide valuable input to future charge radius evaluations.

\begin{acknowledgments}

This work was funded by a NIST Grant Award Number 70NANB20H87 and by a National
Science Foundation Award Number 2309273. E.T.\ gratefully acknowledges the kind hospitality of the research groups at the National Institute for Fusion Science and the University of Electro-Communications during the course of this work.

\end{acknowledgments}

\appendix

\section{\label{app:smooth}Uncertainty of transition-energy differences}

Contributions to transition energies and their uncertainties can be approximated by polynomial fits of the form (e.g., see Fig.~\ref{fig:smooth_function})
\begin{equation}
P_{n}(Z)=a_{n}Z^{n}+a_{n-1}Z^{n-1}+a_{n-2}Z^{n-2}+\cdots\,,\label{eq:polyfit}
\end{equation}
which may be interpreted as a leading term $a_{n}Z^{n}$ supplemented
by correction terms in a $1/Z$ expansion.

In Sec.~\ref{sec:relativistic_theory}, we argue that when the error in the transition energy is a smooth, systematic function of $Z$ (rather than random fluctuations), the uncertainty in the transition-energy difference is obtained by subtraction,
\begin{eqnarray*}
\Delta[E(Z)-E(Z')] & = & |\Delta E(Z)-\Delta E(Z')|\\
 & \approx & |(Z-Z')\Delta E'(Z)|\,,
\end{eqnarray*}
where $\Delta E(Z)$ is the uncertainty in the transition energy $E(Z)$, and $\Delta E'(Z)=d[\Delta E(Z)]/dZ$. To illustrate this, suppose the uncertainty function $\Delta E(Z)$ of a particular contribution is approximated by the leading
term $a_{n}Z^{n}$ of the polynomial fit $P_{n}(Z)$ in Eq.~(\ref{eq:polyfit}). In this case,
\begin{equation*}
\frac{\Delta[E(Z)-E(Z')]}{\Delta E(Z)}\approx\frac{n}{Z}|Z-Z'|\,.
\end{equation*}
Thus, the uncertainty in the Lu-Yb transition-energy difference
is naturally suppressed relative to the uncertainty in the corresponding transition-energy contribution by a factor $n/Z$.

The dominant omitted terms in the QED sector involve the treatment of vertex corrections in the contributions QED(val-x), QED(core), and QED(2-body)~\cite{Gillaspy2013,blundell_calculation_2009}, which scale approximately as $Z^3$. Other errors in the QED sector associated with QED(2-loop) and WK(val) scale with higher powers of $Z$, as do errors in the treatment of retardation in ``1body-ret'' \cite{Gillaspy2013}. Our total estimated uncertainty in the $Z$ range of Fig.~\ref{fig:smooth_function} is found to scale approximately as $a_n Z^n$ with an effective index $n=n_{\text{eff}}\approx 3.5$. This is consistent with the suppression factor $n_{\text{eff}}/Z \approx 1/20$ found in this case (see Table~\ref{tab:V}).

Similarly, the fractional uncertainties in the transition energy $E(Z)$ and in the transition-energy
difference $E(Z)-E(Z')$ are, respectively,
\[
f_{E}=\frac{\Delta E(Z)}{E(Z)}\,, \qquad
f_{\text{diff}}\approx\frac{\Delta E'(Z)}{E'(Z)}\,.
\]
If $E(Z)$ and $\Delta E(Z)$ are represented by leading-order terms $a_{m}Z^{m}$
and $a_{n}Z^{n}$, respectively, then
\[
\frac{f_{\text{diff}}}{f_{E}}\approx\frac{n}{m}\,,
\]
which is of order unity. Hence, the fractional uncertainty in the transition-energy
difference is generally comparable to that of the corresponding transition-energy contribution.

\section{Radii Optimization Procedure \label{app:optimization}}
Given a set of (potentially correlated) measurements with normally distributed errors, non-linear generalized least squares provides the optimal parameters and their uncertainties \cite{kariya_generalized_2004}. For a set of $N$ measurements with mean values $\vec{\mu}$ and covariance matrix $\Sigma$, the optimal set of radii ($\vec{R}$) are those which minimize

\begin{equation}
    \label{eq:GLS}
    \begin{bmatrix}
        \mu_1 - Q_1(\vec{R}) \\ \dots \\ \mu_N - Q_N(\vec{R})
    \end{bmatrix}^T
    \Sigma^{-1}
    \begin{bmatrix}
        \mu_1 - Q_1(\vec{R}) \\ \dots \\ \mu_N - Q_N(\vec{R})
    \end{bmatrix},
\end{equation}
where $Q_i(\vec{R})$ is the quantity determined by the $i$th measurement. For example, if $\vec{R}=\begin{bmatrix}R_{^{175}\text{Lu}} & R_{^{174}\text{Yb}} & \end{bmatrix}^T$, then $Q(\vec{R})$ of an absolute measurement of the radius of $^{174}\text{Yb}$ would simply be $Q(\vec{R}) = \left(\vec{R}\right)_2 = R_{^{174}\text{Yb}}$.

The strong correlation between nuclear charge radii from optical isotope shift measurements can be handled explicitly using this formalism. To first order, frequency shifts between isotopes $A$ and $A'$ can be calculated from

\begin{equation}
    \label{eq:OIS}
    \delta \nu^{AA'} \approx F\delta \langle r^2 \rangle^{AA'} + K\delta\mu^{AA'},
\end{equation}

where $F$ is the field shift coefficient, $K$ is the mass shift coefficient, and $\delta\mu^{AA'}$ is the difference in inverse mass. Rearranging to solve for $\delta \langle r^2 \rangle ^{AA'}\equiv \langle r^2 \rangle^A - \langle r^2 \rangle^{A'}$, we have
\begin{equation}
    \delta\langle r^2 \rangle^{AA'} = \frac{\delta \nu^{AA'} - K\delta\mu^{AA'}}{F} .
\end{equation}

In general, the uncertainty in $\delta\langle r^2 \rangle^{AA'}$ is dominated by theoretical uncertainties in the field and mass shift coefficients. In the case of heavy elements (where the mass shift is much smaller than the field shift) this often means that the uncertainty of OIS radius differences comes almost entirely from the uncertainty in the shared scaling of the field shift coefficient. Indeed, both \cite{GBK98} and \cite{kawasaki_isotopeshift_2024} recommend common scaling uncertainties of 10\% and 20\%, respectively. This common uncertainty factor $s$ can be encoded in the covariance matrix of a set of $N$ nuclear charge radius isotope shifts from optical measurements $\delta\langle r^2 \rangle$ as
\begin{equation}
    \Sigma_{\text{OIS}} = s^2 \begin{bmatrix}
        \delta \langle r^2 \rangle_1 ^2 & \dots & \rho^2 \delta\langle r^2 \rangle_1\delta\langle r^2 \rangle_N \\
        \vdots &   \ddots &\vdots \\
        \rho^2 \delta\langle r^2 \rangle_1\delta\langle r^2 \rangle_N  & \dots & \delta\langle r^2\rangle_N^2
        
    \end{bmatrix}
\end{equation}

where the correlation $\rho$ is close to 1 and with the $\mu\,-\,Q(\vec{R})$ vector from equation \ref{eq:GLS} given by

\begin{equation}
    \vec{\mu}_\text{OIS}-\vec{Q}_\text{OIS}(\vec{R}) = 
    \begin{bmatrix}
    \left(R_1^2 - R_2^2 \right)-\delta\langle r^2\rangle^{A_1A_2} \\
    \vdots \\
    \left(R_1^2 - R_N^2 \right)-\delta\langle r^2\rangle^{A_1A_N}
    \end{bmatrix}.
\end{equation}

The highly charged ion measurements in this work (Eqs. \ref{eq:nalike_r} and \ref{eq:mglike_r}) lead to

\begin{equation}
    \vec{\mu}_\text{HCI} - \vec{Q}_\text{HCI}(\vec{R}) = 
    \begin{bmatrix}
    0.354~\text{fm} - \left(
     \bar{R}_\text{Lu} - \frac{S^{\text{Na}}_\text{Yb}}{{S^\text{Na}_\text{Lu}}} \bar{R}_\text{Yb} \right) 
    \\
    0.547~\text{fm} - \left(\bar{R}_\text{Lu} - \frac{S^{\text{Mg}}_\text{Yb}}{S^\text{Mg}_\text{Lu}} \bar{R}_\text{Yb}\right)
    \end{bmatrix}
\end{equation}
and
\begin{equation}
    \begin{aligned}
    \Sigma_\text{HCI} &= \begin{bmatrix}
        \left(\Delta_\text{Na}\right)^2 & \rho\Delta_\text{Na}\Delta_\text{Mg} \\
        \rho\Delta_\text{Na}\Delta_\text{Mg} & (\Delta_\text{Mg})^2
    \end{bmatrix}\\
    &= \begin{bmatrix}
        143.8 & 21.0 \\
        21.0 & 155.0
    \end{bmatrix} \text{fm}^2.
    \end{aligned}
\end{equation}
Therefore, an absolute measurement for Yb ($R_\text{Yb} = \mu_\text{Yb} \pm \Delta_\text{Yb} $), optical nuclear charge radii isotope shifts, and our HCI constraints determine an optimal set of parameters from the minimization of
\begin{equation}
\begin{aligned}
    &\vec{R}_{\text{opt}} =
    \argmin_{\vec{R}} \Biggl\{ \left(\frac{\mu_\text{Yb}-R_\text{Yb}}{\Delta_\text{Yb}}\right)^2+ \\
    &\left(\vec{\mu}_\text{OIS}-\vec{Q}_\text{OIS}(\vec{R})\right)^T \Sigma_\text{OIS}^{-1} \left(\vec{\mu}_\text{OIS}-\vec{Q}_\text{OIS}(\vec{R})\right) + \\
    &
    \left(\vec{\mu}_\text{HCI}-\vec{Q}_\text{HCI}(\vec{R})\right)^T \Sigma_\text{HCI}^{-1} \left(\vec{\mu}_\text{HCI}-\vec{Q}_\text{HCI}(\vec{R})\right) \biggr \} .\\
\end{aligned}
\end{equation}

\bibliography{references}

\begin{thebibliography}{67}%
\makeatletter
\providecommand \@ifxundefined [1]{%
 \@ifx{#1\undefined}
}%
\providecommand \@ifnum [1]{%
 \ifnum #1\expandafter \@firstoftwo
 \else \expandafter \@secondoftwo
 \fi
}%
\providecommand \@ifx [1]{%
 \ifx #1\expandafter \@firstoftwo
 \else \expandafter \@secondoftwo
 \fi
}%
\providecommand \natexlab [1]{#1}%
\providecommand \enquote  [1]{``#1''}%
\providecommand \bibnamefont  [1]{#1}%
\providecommand \bibfnamefont [1]{#1}%
\providecommand \citenamefont [1]{#1}%
\providecommand \href@noop [0]{\@secondoftwo}%
\providecommand \href [0]{\begingroup \@sanitize@url \@href}%
\providecommand \@href[1]{\@@startlink{#1}\@@href}%
\providecommand \@@href[1]{\endgroup#1\@@endlink}%
\providecommand \@sanitize@url [0]{\catcode `\\12\catcode `\$12\catcode `\&12\catcode `\#12\catcode `\^12\catcode `\_12\catcode `\%12\relax}%
\providecommand \@@startlink[1]{}%
\providecommand \@@endlink[0]{}%
\providecommand \url  [0]{\begingroup\@sanitize@url \@url }%
\providecommand \@url [1]{\endgroup\@href {#1}{\urlprefix }}%
\providecommand \urlprefix  [0]{URL }%
\providecommand \Eprint [0]{\href }%
\providecommand \doibase [0]{https://doi.org/}%
\providecommand \selectlanguage [0]{\@gobble}%
\providecommand \bibinfo  [0]{\@secondoftwo}%
\providecommand \bibfield  [0]{\@secondoftwo}%
\providecommand \translation [1]{[#1]}%
\providecommand \BibitemOpen [0]{}%
\providecommand \bibitemStop [0]{}%
\providecommand \bibitemNoStop [0]{.\EOS\space}%
\providecommand \EOS [0]{\spacefactor3000\relax}%
\providecommand \BibitemShut  [1]{\csname bibitem#1\endcsname}%
\let\auto@bib@innerbib\@empty
\bibitem [{\citenamefont {Silwal}\ \emph {et~al.}(2018)\citenamefont {Silwal}, \citenamefont {Lapierre}, \citenamefont {Gillaspy}, \citenamefont {Dreiling}, \citenamefont {Blundell}, \citenamefont {{Dipti}}, \citenamefont {Borovik}, \citenamefont {Gwinner}, \citenamefont {Villari}, \citenamefont {Ralchenko},\ and\ \citenamefont {Takacs}}]{silwal_measuring_2018}%
  \BibitemOpen
  \bibfield  {author} {\bibinfo {author} {\bibfnamefont {R.}~\bibnamefont {Silwal}}, \bibinfo {author} {\bibfnamefont {A.}~\bibnamefont {Lapierre}}, \bibinfo {author} {\bibfnamefont {J.~D.}\ \bibnamefont {Gillaspy}}, \bibinfo {author} {\bibfnamefont {J.~M.}\ \bibnamefont {Dreiling}}, \bibinfo {author} {\bibfnamefont {S.~A.}\ \bibnamefont {Blundell}}, \bibinfo {author} {\bibnamefont {{Dipti}}}, \bibinfo {author} {\bibfnamefont {A.}~\bibnamefont {Borovik}}, \bibinfo {author} {\bibfnamefont {G.}~\bibnamefont {Gwinner}}, \bibinfo {author} {\bibfnamefont {A.~C.~C.}\ \bibnamefont {Villari}}, \bibinfo {author} {\bibfnamefont {Y.}~\bibnamefont {Ralchenko}},\ and\ \bibinfo {author} {\bibfnamefont {E.}~\bibnamefont {Takacs}},\ }\bibfield  {title} {\bibinfo {title} {Measuring the difference in nuclear charge radius of {Xe} isotopes by {EUV} spectroscopy of highly charged {Na}-like ions},\ }\href {https://doi.org/10.1103/PhysRevA.98.052502} {\bibfield  {journal} {\bibinfo  {journal} {Phys. Rev. A}\ }\textbf {\bibinfo
  {volume} {98}},\ \bibinfo {pages} {052502} (\bibinfo {year} {2018})}\BibitemShut {NoStop}%
\bibitem [{\citenamefont {Silwal}\ \emph {et~al.}(2020)\citenamefont {Silwal}, \citenamefont {Lapierre}, \citenamefont {Gillaspy}, \citenamefont {Dreiling}, \citenamefont {Blundell}, \citenamefont {{Dipti}}, \citenamefont {Borovik}, \citenamefont {Gwinner}, \citenamefont {Villari}, \citenamefont {Ralchenko},\ and\ \citenamefont {Takacs}}]{silwal_determination_2020}%
  \BibitemOpen
  \bibfield  {author} {\bibinfo {author} {\bibfnamefont {R.}~\bibnamefont {Silwal}}, \bibinfo {author} {\bibfnamefont {A.}~\bibnamefont {Lapierre}}, \bibinfo {author} {\bibfnamefont {J.~D.}\ \bibnamefont {Gillaspy}}, \bibinfo {author} {\bibfnamefont {J.~M.}\ \bibnamefont {Dreiling}}, \bibinfo {author} {\bibfnamefont {S.~A.}\ \bibnamefont {Blundell}}, \bibinfo {author} {\bibnamefont {{Dipti}}}, \bibinfo {author} {\bibfnamefont {A.}~\bibnamefont {Borovik}}, \bibinfo {author} {\bibfnamefont {G.}~\bibnamefont {Gwinner}}, \bibinfo {author} {\bibfnamefont {A.~C.~C.}\ \bibnamefont {Villari}}, \bibinfo {author} {\bibfnamefont {Y.}~\bibnamefont {Ralchenko}},\ and\ \bibinfo {author} {\bibfnamefont {E.}~\bibnamefont {Takacs}},\ }\bibfield  {title} {\bibinfo {title} {Determination of the isotopic change in nuclear charge radius from extreme-ultraviolet spectroscopy of highly charged ions of {Xe}},\ }\href {https://doi.org/10.1103/PhysRevA.101.062512} {\bibfield  {journal} {\bibinfo  {journal} {Phys. Rev. A}\ }\textbf
  {\bibinfo {volume} {101}},\ \bibinfo {pages} {062512} (\bibinfo {year} {2020})}\BibitemShut {NoStop}%
\bibitem [{\citenamefont {Hosier}\ \emph {et~al.}(2024)\citenamefont {Hosier}, \citenamefont {{Dipti}}, \citenamefont {Blundell}, \citenamefont {Silwal}, \citenamefont {Lapierre}, \citenamefont {Gillaspy}, \citenamefont {Gwinner}, \citenamefont {Tan}, \citenamefont {Kwiatkowski}, \citenamefont {Wang}, \citenamefont {Staiger}, \citenamefont {Villari}, \citenamefont {Ralchenko},\ and\ \citenamefont {Takacs}}]{hosier_absolute_2024}%
  \BibitemOpen
  \bibfield  {author} {\bibinfo {author} {\bibfnamefont {A.}~\bibnamefont {Hosier}}, \bibinfo {author} {\bibnamefont {{Dipti}}}, \bibinfo {author} {\bibfnamefont {S.~A.}\ \bibnamefont {Blundell}}, \bibinfo {author} {\bibfnamefont {R.}~\bibnamefont {Silwal}}, \bibinfo {author} {\bibfnamefont {A.}~\bibnamefont {Lapierre}}, \bibinfo {author} {\bibfnamefont {J.~D.}\ \bibnamefont {Gillaspy}}, \bibinfo {author} {\bibfnamefont {G.}~\bibnamefont {Gwinner}}, \bibinfo {author} {\bibfnamefont {J.~N.}\ \bibnamefont {Tan}}, \bibinfo {author} {\bibfnamefont {A.~A.}\ \bibnamefont {Kwiatkowski}}, \bibinfo {author} {\bibfnamefont {Y.}~\bibnamefont {Wang}}, \bibinfo {author} {\bibfnamefont {H.}~\bibnamefont {Staiger}}, \bibinfo {author} {\bibfnamefont {A.~C.~C.}\ \bibnamefont {Villari}}, \bibinfo {author} {\bibfnamefont {Y.}~\bibnamefont {Ralchenko}},\ and\ \bibinfo {author} {\bibfnamefont {E.}~\bibnamefont {Takacs}},\ }\bibfield  {title} {\bibinfo {title} {Absolute nuclear charge radius by {{Na-like}} spectral line separation
  in high-{{Z}} elements},\ }\href {https://doi.org/10.1088/1361-6455/ad717b} {\bibfield  {journal} {\bibinfo  {journal} {J. Phys. B: At. Mol. Opt. Phys.}\ }\textbf {\bibinfo {volume} {57}},\ \bibinfo {pages} {195001} (\bibinfo {year} {2024})}\BibitemShut {NoStop}%
\bibitem [{\citenamefont {Hosier}\ \emph {et~al.}(2025)\citenamefont {Hosier}, \citenamefont {Dipti}, \citenamefont {Blundell}, \citenamefont {Lapierre}, \citenamefont {Silwal}, \citenamefont {Gwinner}, \citenamefont {Tan}, \citenamefont {Naing}, \citenamefont {Gillaspy}, \citenamefont {Yang} \emph {et~al.}}]{Hosier2025}%
  \BibitemOpen
  \bibfield  {author} {\bibinfo {author} {\bibfnamefont {A.}~\bibnamefont {Hosier}}, \bibinfo {author} {\bibnamefont {Dipti}}, \bibinfo {author} {\bibfnamefont {S.~A.}\ \bibnamefont {Blundell}}, \bibinfo {author} {\bibfnamefont {A.}~\bibnamefont {Lapierre}}, \bibinfo {author} {\bibfnamefont {R.}~\bibnamefont {Silwal}}, \bibinfo {author} {\bibfnamefont {G.}~\bibnamefont {Gwinner}}, \bibinfo {author} {\bibfnamefont {J.~N.}\ \bibnamefont {Tan}}, \bibinfo {author} {\bibfnamefont {A.}~\bibnamefont {Naing}}, \bibinfo {author} {\bibfnamefont {J.~D.}\ \bibnamefont {Gillaspy}}, \bibinfo {author} {\bibfnamefont {Y.}~\bibnamefont {Yang}}, \emph {et~al.},\ }\bibfield  {title} {\bibinfo {title} {Determination of nuclear charge radius by extreme-ultraviolet spectroscopy of {Na}-like ions},\ }\href {https://doi.org/10.1103/PhysRevResearch.7.L012024} {\bibfield  {journal} {\bibinfo  {journal} {Phys. Rev. Res.}\ }\textbf {\bibinfo {volume} {7}},\ \bibinfo {pages} {L012024} (\bibinfo {year} {2025})}\BibitemShut {NoStop}%
\bibitem [{\citenamefont {Takacs}\ \emph {et~al.}(2025)\citenamefont {Takacs}, \citenamefont {Staiger}, \citenamefont {Blundell}, \citenamefont {Kimura}, \citenamefont {Sakaue}, \citenamefont {Suzuki}, \citenamefont {Dipti}, \citenamefont {Angeli}, \citenamefont {Faiyaz}, \citenamefont {{{Yu.} Ralchenko}}, \citenamefont {Murakami}, \citenamefont {Kato}, \citenamefont {Ruiz}, \citenamefont {Nagai}, \citenamefont {Takaoka} \emph {et~al.}}]{companionTakacs2025}%
  \BibitemOpen
  \bibfield  {author} {\bibinfo {author} {\bibfnamefont {E.}~\bibnamefont {Takacs}}, \bibinfo {author} {\bibfnamefont {H.}~\bibnamefont {Staiger}}, \bibinfo {author} {\bibfnamefont {S.~A.}\ \bibnamefont {Blundell}}, \bibinfo {author} {\bibfnamefont {N.}~\bibnamefont {Kimura}}, \bibinfo {author} {\bibfnamefont {H.~A.}\ \bibnamefont {Sakaue}}, \bibinfo {author} {\bibfnamefont {C.}~\bibnamefont {Suzuki}}, \bibinfo {author} {\bibnamefont {Dipti}}, \bibinfo {author} {\bibfnamefont {I.}~\bibnamefont {Angeli}}, \bibinfo {author} {\bibfnamefont {C.~A.}\ \bibnamefont {Faiyaz}}, \bibinfo {author} {\bibnamefont {{{Yu.} Ralchenko}}}, \bibinfo {author} {\bibfnamefont {I.}~\bibnamefont {Murakami}}, \bibinfo {author} {\bibfnamefont {D.}~\bibnamefont {Kato}}, \bibinfo {author} {\bibfnamefont {R.~G.}\ \bibnamefont {Ruiz}}, \bibinfo {author} {\bibfnamefont {Y.}~\bibnamefont {Nagai}}, \bibinfo {author} {\bibfnamefont {R.}~\bibnamefont {Takaoka}}, \emph {et~al.},\ }\bibfield  {title} {\bibinfo {title} {{Puzzling Isotonic
  Odd-Even Staggering of Charge Radii in Deformed Rare Earth Nuclei}},\ }\href@noop {} {\bibfield  {journal} {\bibinfo  {journal} {arXiv preprint arXiv:2511.19395}\ } (\bibinfo {year} {2025})},\ \Eprint {https://arxiv.org/abs/2511.19395} {arXiv:2511.19395 [physics.atom-ph]} \BibitemShut {NoStop}%
\bibitem [{\citenamefont {Nakamura}\ \emph {et~al.}(1997)\citenamefont {Nakamura}, \citenamefont {Asada}, \citenamefont {Currell}, \citenamefont {Fukami}, \citenamefont {Hirayama}, \citenamefont {Motohashi}, \citenamefont {Nagata}, \citenamefont {Nojikawa}, \citenamefont {Ohtani}, \citenamefont {Okazaki}, \citenamefont {Sakurai}, \citenamefont {Shiraishi}, \citenamefont {Tsurubuchi},\ and\ \citenamefont {Watanabe}}]{Nakamura1997}%
  \BibitemOpen
  \bibfield  {author} {\bibinfo {author} {\bibfnamefont {N.}~\bibnamefont {Nakamura}}, \bibinfo {author} {\bibfnamefont {J.}~\bibnamefont {Asada}}, \bibinfo {author} {\bibfnamefont {F.~J.}\ \bibnamefont {Currell}}, \bibinfo {author} {\bibfnamefont {T.}~\bibnamefont {Fukami}}, \bibinfo {author} {\bibfnamefont {T.}~\bibnamefont {Hirayama}}, \bibinfo {author} {\bibfnamefont {K.}~\bibnamefont {Motohashi}}, \bibinfo {author} {\bibfnamefont {T.}~\bibnamefont {Nagata}}, \bibinfo {author} {\bibfnamefont {E.}~\bibnamefont {Nojikawa}}, \bibinfo {author} {\bibfnamefont {S.}~\bibnamefont {Ohtani}}, \bibinfo {author} {\bibfnamefont {K.}~\bibnamefont {Okazaki}}, \bibinfo {author} {\bibfnamefont {M.}~\bibnamefont {Sakurai}}, \bibinfo {author} {\bibfnamefont {H.}~\bibnamefont {Shiraishi}}, \bibinfo {author} {\bibfnamefont {S.}~\bibnamefont {Tsurubuchi}},\ and\ \bibinfo {author} {\bibfnamefont {H.}~\bibnamefont {Watanabe}},\ }\bibfield  {title} {\bibinfo {title} {An overview of the {T}okyo electron beam ion trap},\ }\href
  {https://doi.org/10.1088/0031-8949/1997/T73/119} {\bibfield  {journal} {\bibinfo  {journal} {Phys. Scr.}\ }\textbf {\bibinfo {volume} {1997}},\ \bibinfo {pages} {362} (\bibinfo {year} {1997})}\BibitemShut {NoStop}%
\bibitem [{\citenamefont {Bender}\ \emph {et~al.}(2003)\citenamefont {Bender}, \citenamefont {Heenen},\ and\ \citenamefont {Reinhard}}]{Bender2003}%
  \BibitemOpen
  \bibfield  {author} {\bibinfo {author} {\bibfnamefont {M.}~\bibnamefont {Bender}}, \bibinfo {author} {\bibfnamefont {P.-H.}\ \bibnamefont {Heenen}},\ and\ \bibinfo {author} {\bibfnamefont {P.-G.}\ \bibnamefont {Reinhard}},\ }\bibfield  {title} {\bibinfo {title} {Self-consistent mean-field models for nuclear structure},\ }\href {https://doi.org/10.1103/RevModPhys.75.121} {\bibfield  {journal} {\bibinfo  {journal} {Rev. Mod. Phys.}\ }\textbf {\bibinfo {volume} {75}},\ \bibinfo {pages} {121} (\bibinfo {year} {2003})}\BibitemShut {NoStop}%
\bibitem [{\citenamefont {Fayans}(1998)}]{Fayans1998}%
  \BibitemOpen
  \bibfield  {author} {\bibinfo {author} {\bibfnamefont {S.~A.}\ \bibnamefont {Fayans}},\ }\bibfield  {title} {\bibinfo {title} {Towards a universal nuclear density functional},\ }\href {https://doi.org/10.1134/1.567841} {\bibfield  {journal} {\bibinfo  {journal} {J. Exp. Theor. Phys. Lett.}\ }\textbf {\bibinfo {volume} {68}},\ \bibinfo {pages} {169} (\bibinfo {year} {1998})}\BibitemShut {NoStop}%
\bibitem [{\citenamefont {Fayans}\ \emph {et~al.}(2000)\citenamefont {Fayans}, \citenamefont {Tolokonnikov}, \citenamefont {Trykov},\ and\ \citenamefont {Zawischa}}]{Fayans2000}%
  \BibitemOpen
  \bibfield  {author} {\bibinfo {author} {\bibfnamefont {S.~A.}\ \bibnamefont {Fayans}}, \bibinfo {author} {\bibfnamefont {S.~V.}\ \bibnamefont {Tolokonnikov}}, \bibinfo {author} {\bibfnamefont {E.~L.}\ \bibnamefont {Trykov}},\ and\ \bibinfo {author} {\bibfnamefont {D.}~\bibnamefont {Zawischa}},\ }\bibfield  {title} {\bibinfo {title} {Nuclear isotope shifts within the local energy-density functional approach},\ }\href {https://doi.org/10.1016/S0375-9474(00)00192-5} {\bibfield  {journal} {\bibinfo  {journal} {Nucl. Phys. A}\ }\textbf {\bibinfo {volume} {676}},\ \bibinfo {pages} {49} (\bibinfo {year} {2000})}\BibitemShut {NoStop}%
\bibitem [{\citenamefont {Sasanuma}(1979)}]{Sasanuma1979}%
  \BibitemOpen
  \bibfield  {author} {\bibinfo {author} {\bibfnamefont {T.}~\bibnamefont {Sasanuma}},\ }\emph {\bibinfo {title} {{Electron Scattering from Deformed Heavy Nuclei}}},\ \href {https://dspace.mit.edu/handle/1721.1/122197} {\bibinfo {type} {Ph.{D}. thesis}},\ \bibinfo  {school} {Massachusetts Institute of Technology} (\bibinfo {year} {1979})\BibitemShut {NoStop}%
\bibitem [{\citenamefont {Suzuki}(1968)}]{Su68}%
  \BibitemOpen
  \bibfield  {author} {\bibinfo {author} {\bibfnamefont {N.~N.}\ \bibnamefont {Suzuki}},\ }\href@noop {} {Ph.D. thesis},\ \bibinfo  {school} {Carnegie Mellon University} (\bibinfo {year} {1968})\BibitemShut {NoStop}%
\bibitem [{\citenamefont {Engfer}\ \emph {et~al.}(1974)\citenamefont {Engfer}, \citenamefont {Schneuwly}, \citenamefont {Vuilleumier}, \citenamefont {Walter},\ and\ \citenamefont {Zehnder}}]{ESV74}%
  \BibitemOpen
  \bibfield  {author} {\bibinfo {author} {\bibfnamefont {R.}~\bibnamefont {Engfer}}, \bibinfo {author} {\bibfnamefont {H.}~\bibnamefont {Schneuwly}}, \bibinfo {author} {\bibfnamefont {J.}~\bibnamefont {Vuilleumier}}, \bibinfo {author} {\bibfnamefont {H.}~\bibnamefont {Walter}},\ and\ \bibinfo {author} {\bibfnamefont {A.}~\bibnamefont {Zehnder}},\ }\bibfield  {title} {\bibinfo {title} {Charge-distribution parameters, isotope shifts, isomer shifts, and magnetic hyperfine constants from muonic atoms},\ }\href {https://doi.org/https://doi.org/10.1016/S0092-640X(74)80003-3} {\bibfield  {journal} {\bibinfo  {journal} {At. Data Nucl. Data Tables}\ }\textbf {\bibinfo {volume} {14}},\ \bibinfo {pages} {509} (\bibinfo {year} {1974})}\BibitemShut {NoStop}%
\bibitem [{\citenamefont {Angeli}\ and\ \citenamefont {Marinova}(2013)}]{Angeli13}%
  \BibitemOpen
  \bibfield  {author} {\bibinfo {author} {\bibfnamefont {I.}~\bibnamefont {Angeli}}\ and\ \bibinfo {author} {\bibfnamefont {K.~P.}\ \bibnamefont {Marinova}},\ }\bibfield  {title} {\bibinfo {title} {Table of experimental nuclear ground state charge radii: An update},\ }\href {https://doi.org/https://doi.org/10.1016/j.adt.2011.12.006} {\bibfield  {journal} {\bibinfo  {journal} {At. Data Nucl. Data Tables}\ }\textbf {\bibinfo {volume} {99}},\ \bibinfo {pages} {69} (\bibinfo {year} {2013})}\BibitemShut {NoStop}%
\bibitem [{\citenamefont {Fricke}\ and\ \citenamefont {Heilig}(2004)}]{fricke_nuclear_2004}%
  \BibitemOpen
  \bibfield  {author} {\bibinfo {author} {\bibfnamefont {G.}~\bibnamefont {Fricke}}\ and\ \bibinfo {author} {\bibfnamefont {K.}~\bibnamefont {Heilig}},\ }\href {https://doi.org/10.1007/b87879} {\emph {\bibinfo {title} {Nuclear Charge Radii}}},\ edited by\ \bibinfo {editor} {\bibfnamefont {H.}~\bibnamefont {Schopper}}\ (\bibinfo  {publisher} {Springer Materials},\ \bibinfo {year} {2004})\BibitemShut {NoStop}%
\bibitem [{\citenamefont {Zehnder}\ \emph {et~al.}(1975)\citenamefont {Zehnder}, \citenamefont {Boehm}, \citenamefont {Dey}, \citenamefont {Engfer}, \citenamefont {Walter},\ and\ \citenamefont {Vuilleumier}}]{zehnder_charge_1975}%
  \BibitemOpen
  \bibfield  {author} {\bibinfo {author} {\bibfnamefont {A.}~\bibnamefont {Zehnder}}, \bibinfo {author} {\bibfnamefont {F.}~\bibnamefont {Boehm}}, \bibinfo {author} {\bibfnamefont {W.}~\bibnamefont {Dey}}, \bibinfo {author} {\bibfnamefont {R.}~\bibnamefont {Engfer}}, \bibinfo {author} {\bibfnamefont {H.}~\bibnamefont {Walter}},\ and\ \bibinfo {author} {\bibfnamefont {J.}~\bibnamefont {Vuilleumier}},\ }\bibfield  {title} {\bibinfo {title} {Charge parameters, isotope shifts, quadrupole moments, and nuclear excitation in muonic \textsuperscript{170–174,176}{Yb}},\ }\href {https://doi.org/https://doi.org/10.1016/0375-9474(75)90219-5} {\bibfield  {journal} {\bibinfo  {journal} {Nucl. Phys. A}\ }\textbf {\bibinfo {volume} {254}},\ \bibinfo {pages} {315} (\bibinfo {year} {1975})}\BibitemShut {NoStop}%
\bibitem [{\citenamefont {Angeli}(1999)}]{Angeli1999}%
  \BibitemOpen
  \bibfield  {author} {\bibinfo {author} {\bibfnamefont {I.}~\bibnamefont {Angeli}},\ }\href@noop {} {\emph {\bibinfo {title} {{Table of Nuclear Root Mean Square Charge Radii}}}},\ \bibinfo {type} {Tech. Rep.}\ \bibinfo {number} {INDC(HUN)-033}\ (\bibinfo  {institution} {IAEA Nuclear Data Section},\ \bibinfo {year} {1999})\ \bibinfo {note} {available at \url{http://www-nds.iaea.or.at/indc-sel.html}}\BibitemShut {NoStop}%
\bibitem [{\citenamefont {Adler}(1975)}]{adler_study_1975}%
  \BibitemOpen
  \bibfield  {author} {\bibinfo {author} {\bibfnamefont {D.~T.}\ \bibnamefont {Adler}},\ }\emph {\bibinfo {title} {A study of the muonic x-rays from separated isotopes of Ytterbium}},\ \href@noop {} {Ph.D. thesis},\ \bibinfo  {school} {Carnegie Mellon University} (\bibinfo {year} {1975})\BibitemShut {NoStop}%
\bibitem [{\citenamefont {Bernhardt}(1992)}]{bernhardt_1992}%
  \BibitemOpen
  \bibfield  {author} {\bibinfo {author} {\bibfnamefont {C.}~\bibnamefont {Bernhardt}},\ }\href@noop {} {Ph.D. thesis},\ \bibinfo  {school} {Universitat Mainz} (\bibinfo {year} {1992})\BibitemShut {NoStop}%
\bibitem [{\citenamefont {Cooper}\ \emph {et~al.}(1976)\citenamefont {Cooper}, \citenamefont {Bertozzi}, \citenamefont {Heisemberg}, \citenamefont {Kowalski}, \citenamefont {Turchinetz}, \citenamefont {Williamson}, \citenamefont {Cardman}, \citenamefont {Fivozinsky}, \citenamefont {Lightbody},\ and\ \citenamefont {Penner}}]{cooper_shapes_1976}%
  \BibitemOpen
  \bibfield  {author} {\bibinfo {author} {\bibfnamefont {T.}~\bibnamefont {Cooper}}, \bibinfo {author} {\bibfnamefont {W.}~\bibnamefont {Bertozzi}}, \bibinfo {author} {\bibfnamefont {J.}~\bibnamefont {Heisemberg}}, \bibinfo {author} {\bibfnamefont {S.}~\bibnamefont {Kowalski}}, \bibinfo {author} {\bibfnamefont {W.}~\bibnamefont {Turchinetz}}, \bibinfo {author} {\bibfnamefont {C.}~\bibnamefont {Williamson}}, \bibinfo {author} {\bibfnamefont {L.}~\bibnamefont {Cardman}}, \bibinfo {author} {\bibfnamefont {S.}~\bibnamefont {Fivozinsky}}, \bibinfo {author} {\bibfnamefont {J.}~\bibnamefont {Lightbody}},\ and\ \bibinfo {author} {\bibfnamefont {S.}~\bibnamefont {Penner}},\ }\bibfield  {title} {\bibinfo {title} {Shapes of deformed nuclei as determined by electron scattering: $^{152}\mathrm{Sm}$, $^{154}\mathrm{Sm}$, $^{166}\mathrm{Er}$, $^{176}\mathrm{Yb}$, $^{232}\mathrm{Th}$, and $^{238}\mathrm{U}$},\ }\href {https://doi.org/10.1103/PhysRevC.13.1083} {\bibfield  {journal} {\bibinfo  {journal} {Phys. Rev. C}\
  }\textbf {\bibinfo {volume} {13}},\ \bibinfo {pages} {1083} (\bibinfo {year} {1976})}\BibitemShut {NoStop}%
\bibitem [{\citenamefont {Creswell}(1977)}]{creswell_electron_1977}%
  \BibitemOpen
  \bibfield  {author} {\bibinfo {author} {\bibfnamefont {C.~W.}\ \bibnamefont {Creswell}},\ }\emph {\bibinfo {title} {Electron Scattering Studies of $^{166}${Er}, $^{176}${Yb}, And $^{238}${U}}},\ \href {https://hdl.handle.net/1721.1/138549} {Ph.D. thesis},\ \bibinfo  {school} {Massachutes Institute of Technology} (\bibinfo {year} {1977})\BibitemShut {NoStop}%
\bibitem [{\citenamefont {De~Vries}\ and\ \citenamefont {De~Vries}(1987)}]{devries_nuclear_1987}%
  \BibitemOpen
  \bibfield  {author} {\bibinfo {author} {\bibfnamefont {C.~W.}\ \bibnamefont {De~Vries}, \bibfnamefont {H.~De~Jager}}\ and\ \bibinfo {author} {\bibfnamefont {C.}~\bibnamefont {De~Vries}},\ }\bibfield  {title} {\bibinfo {title} {Nuclear charge-density-distribution parameters from elastic electron scattering},\ }\href {https://doi.org/https://doi.org/10.1016/0092-640X(87)90013-1} {\bibfield  {journal} {\bibinfo  {journal} {At. Data Nucl. Data Tables}\ }\textbf {\bibinfo {volume} {36}},\ \bibinfo {pages} {495} (\bibinfo {year} {1987})}\BibitemShut {NoStop}%
\bibitem [{\citenamefont {Currell}\ \emph {et~al.}(1996)\citenamefont {Currell}, \citenamefont {Asada}, \citenamefont {Ishii}, \citenamefont {Minoh}, \citenamefont {Motohashi}, \citenamefont {Nakamura}, \citenamefont {Nishizawa}, \citenamefont {Ohtani}, \citenamefont {Okazaki}, \citenamefont {Sakurai}, \citenamefont {Shiraishi}, \citenamefont {Tsurubuchi},\ and\ \citenamefont {Watanabe}}]{Currell1996}%
  \BibitemOpen
  \bibfield  {author} {\bibinfo {author} {\bibfnamefont {F.~J.}\ \bibnamefont {Currell}}, \bibinfo {author} {\bibfnamefont {J.}~\bibnamefont {Asada}}, \bibinfo {author} {\bibfnamefont {K.}~\bibnamefont {Ishii}}, \bibinfo {author} {\bibfnamefont {A.}~\bibnamefont {Minoh}}, \bibinfo {author} {\bibfnamefont {K.}~\bibnamefont {Motohashi}}, \bibinfo {author} {\bibfnamefont {N.}~\bibnamefont {Nakamura}}, \bibinfo {author} {\bibfnamefont {K.}~\bibnamefont {Nishizawa}}, \bibinfo {author} {\bibfnamefont {S.}~\bibnamefont {Ohtani}}, \bibinfo {author} {\bibfnamefont {K.}~\bibnamefont {Okazaki}}, \bibinfo {author} {\bibfnamefont {M.}~\bibnamefont {Sakurai}}, \bibinfo {author} {\bibfnamefont {H.}~\bibnamefont {Shiraishi}}, \bibinfo {author} {\bibfnamefont {S.}~\bibnamefont {Tsurubuchi}},\ and\ \bibinfo {author} {\bibfnamefont {H.}~\bibnamefont {Watanabe}},\ }\bibfield  {title} {\bibinfo {title} {A new versatile electron-beam ion trap},\ }\href {https://doi.org/https://doi.org/10.1143/JPSJ.65.3186} {\bibfield  {journal}
  {\bibinfo  {journal} {J. Phys. Soc. Jpn.}\ }\textbf {\bibinfo {volume} {65}},\ \bibinfo {pages} {3186} (\bibinfo {year} {1996})}\BibitemShut {NoStop}%
\bibitem [{\citenamefont {Yamada}\ \emph {et~al.}(2006)\citenamefont {Yamada}, \citenamefont {Nagata}, \citenamefont {Nakamura}, \citenamefont {Ohtani}, \citenamefont {Takahashi}, \citenamefont {Tobiyama}, \citenamefont {Tona}, \citenamefont {Watanabe}, \citenamefont {Yoshiyasu}, \citenamefont {Sakurai}, \citenamefont {Kavanagh},\ and\ \citenamefont {Currell}}]{Yamada2006}%
  \BibitemOpen
  \bibfield  {author} {\bibinfo {author} {\bibfnamefont {C.}~\bibnamefont {Yamada}}, \bibinfo {author} {\bibfnamefont {K.}~\bibnamefont {Nagata}}, \bibinfo {author} {\bibfnamefont {N.}~\bibnamefont {Nakamura}}, \bibinfo {author} {\bibfnamefont {S.}~\bibnamefont {Ohtani}}, \bibinfo {author} {\bibfnamefont {S.}~\bibnamefont {Takahashi}}, \bibinfo {author} {\bibfnamefont {T.}~\bibnamefont {Tobiyama}}, \bibinfo {author} {\bibfnamefont {M.}~\bibnamefont {Tona}}, \bibinfo {author} {\bibfnamefont {H.}~\bibnamefont {Watanabe}}, \bibinfo {author} {\bibfnamefont {N.}~\bibnamefont {Yoshiyasu}}, \bibinfo {author} {\bibfnamefont {M.}~\bibnamefont {Sakurai}}, \bibinfo {author} {\bibfnamefont {A.~P.}\ \bibnamefont {Kavanagh}},\ and\ \bibinfo {author} {\bibfnamefont {F.~J.}\ \bibnamefont {Currell}},\ }\bibfield  {title} {\bibinfo {title} {Injection of metallic elements into an electron-beam ion trap using a {K}nudsen cell},\ }\href {https://doi.org/10.1063/1.2216867} {\bibfield  {journal} {\bibinfo  {journal} {Rev. Sci.
  Instrum.}\ }\textbf {\bibinfo {volume} {77}},\ \bibinfo {pages} {066110} (\bibinfo {year} {2006})}\BibitemShut {NoStop}%
\bibitem [{\citenamefont {Nakamura}\ \emph {et~al.}(2000)\citenamefont {Nakamura}, \citenamefont {Kinugawa}, \citenamefont {Shimizu}, \citenamefont {Watanabe}, \citenamefont {Ito}, \citenamefont {Ohtani}, \citenamefont {Yamada}, \citenamefont {Okazaki}, \citenamefont {Sakurai}, \citenamefont {Tarbutt},\ and\ \citenamefont {Silver}}]{Nakamura2000}%
  \BibitemOpen
  \bibfield  {author} {\bibinfo {author} {\bibfnamefont {N.}~\bibnamefont {Nakamura}}, \bibinfo {author} {\bibfnamefont {T.}~\bibnamefont {Kinugawa}}, \bibinfo {author} {\bibfnamefont {H.}~\bibnamefont {Shimizu}}, \bibinfo {author} {\bibfnamefont {H.}~\bibnamefont {Watanabe}}, \bibinfo {author} {\bibfnamefont {S.}~\bibnamefont {Ito}}, \bibinfo {author} {\bibfnamefont {S.}~\bibnamefont {Ohtani}}, \bibinfo {author} {\bibfnamefont {C.}~\bibnamefont {Yamada}}, \bibinfo {author} {\bibfnamefont {K.}~\bibnamefont {Okazaki}}, \bibinfo {author} {\bibfnamefont {M.}~\bibnamefont {Sakurai}}, \bibinfo {author} {\bibfnamefont {M.~R.}\ \bibnamefont {Tarbutt}},\ and\ \bibinfo {author} {\bibfnamefont {J.~D.}\ \bibnamefont {Silver}},\ }\bibfield  {title} {\bibinfo {title} {Injection of various metallic elements into an electron beam ion trap: Techniques needed for systematic investigations of isoelectronic sequences},\ }\href {https://doi.org/10.1063/1.1150260} {\bibfield  {journal} {\bibinfo  {journal} {Rev. Sci. Instrum.}\
  }\textbf {\bibinfo {volume} {71}},\ \bibinfo {pages} {684} (\bibinfo {year} {2000})}\BibitemShut {NoStop}%
\bibitem [{\citenamefont {Kramida}\ and\ \citenamefont {{Buchet-Poulizac}}(2006{\natexlab{a}})}]{kramida_nevii_2006}%
  \BibitemOpen
  \bibfield  {author} {\bibinfo {author} {\bibfnamefont {A.}~\bibnamefont {Kramida}}\ and\ \bibinfo {author} {\bibfnamefont {M.-C.}\ \bibnamefont {{Buchet-Poulizac}}},\ }\bibfield  {title} {\bibinfo {title} {Energy levels and spectral lines of {{Ne VII}}},\ }\href {https://doi.org/10.1140/epjd/e2006-00025-3} {\bibfield  {journal} {\bibinfo  {journal} {Eur. Phys. J. D}\ }\textbf {\bibinfo {volume} {38}},\ \bibinfo {pages} {265} (\bibinfo {year} {2006}{\natexlab{a}})}\BibitemShut {NoStop}%
\bibitem [{\citenamefont {Kramida}\ and\ \citenamefont {{Buchet-Poulizac}}(2006{\natexlab{b}})}]{kramida_neviii_2006}%
  \BibitemOpen
  \bibfield  {author} {\bibinfo {author} {\bibfnamefont {A.}~\bibnamefont {Kramida}}\ and\ \bibinfo {author} {\bibfnamefont {M.-C.}\ \bibnamefont {{Buchet-Poulizac}}},\ }\bibfield  {title} {\bibinfo {title} {Energy levels and spectral lines of {{Ne VIII}}},\ }\href {https://doi.org/10.1140/epjd/e2006-00122-3} {\bibfield  {journal} {\bibinfo  {journal} {Eur. Phys. J. D}\ }\textbf {\bibinfo {volume} {39}},\ \bibinfo {pages} {173} (\bibinfo {year} {2006}{\natexlab{b}})}\BibitemShut {NoStop}%
\bibitem [{\citenamefont {Ralchenko}\ \emph {et~al.}(2008)\citenamefont {Ralchenko}, \citenamefont {Draganic}, \citenamefont {Tan}, \citenamefont {Gillaspy}, \citenamefont {Pomeroy}, \citenamefont {Reader}, \citenamefont {Feldman},\ and\ \citenamefont {Holland}}]{Ralchenko2008}%
  \BibitemOpen
  \bibfield  {author} {\bibinfo {author} {\bibfnamefont {Y.}~\bibnamefont {Ralchenko}}, \bibinfo {author} {\bibfnamefont {I.~N.}\ \bibnamefont {Draganic}}, \bibinfo {author} {\bibfnamefont {J.~N.}\ \bibnamefont {Tan}}, \bibinfo {author} {\bibfnamefont {J.~D.}\ \bibnamefont {Gillaspy}}, \bibinfo {author} {\bibfnamefont {J.~M.}\ \bibnamefont {Pomeroy}}, \bibinfo {author} {\bibfnamefont {J.}~\bibnamefont {Reader}}, \bibinfo {author} {\bibfnamefont {U.}~\bibnamefont {Feldman}},\ and\ \bibinfo {author} {\bibfnamefont {G.~E.}\ \bibnamefont {Holland}},\ }\bibfield  {title} {\bibinfo {title} {{EUV spectra of highly-charged ions W\textsuperscript{54+} - W\textsuperscript{63+} relevant to ITER diagnostics}},\ }\href {https://doi.org/10.1088/0953-4075/41/2/021003} {\bibfield  {journal} {\bibinfo  {journal} {J. Phys. B: At. Mol. Opt. Phys.}\ }\textbf {\bibinfo {volume} {41}},\ \bibinfo {pages} {021003} (\bibinfo {year} {2008})}\BibitemShut {NoStop}%
\bibitem [{\citenamefont {Gillaspy}\ \emph {et~al.}(2009)\citenamefont {Gillaspy}, \citenamefont {Draganić}, \citenamefont {Ralchenko}, \citenamefont {Reader}, \citenamefont {Tan}, \citenamefont {Pomeroy},\ and\ \citenamefont {Brewer}}]{Gillaspy2009}%
  \BibitemOpen
  \bibfield  {author} {\bibinfo {author} {\bibfnamefont {J.~D.}\ \bibnamefont {Gillaspy}}, \bibinfo {author} {\bibfnamefont {I.~N.}\ \bibnamefont {Draganić}}, \bibinfo {author} {\bibfnamefont {Y.}~\bibnamefont {Ralchenko}}, \bibinfo {author} {\bibfnamefont {J.}~\bibnamefont {Reader}}, \bibinfo {author} {\bibfnamefont {J.~N.}\ \bibnamefont {Tan}}, \bibinfo {author} {\bibfnamefont {J.~M.}\ \bibnamefont {Pomeroy}},\ and\ \bibinfo {author} {\bibfnamefont {S.~M.}\ \bibnamefont {Brewer}},\ }\bibfield  {title} {\bibinfo {title} {Measurement of the {D}-line doublet in high-{Z} highly charged sodiumlike ions},\ }\href {https://doi.org/10.1103/PhysRevA.80.010501} {\bibfield  {journal} {\bibinfo  {journal} {Phys. Rev. A}\ }\textbf {\bibinfo {volume} {80}},\ \bibinfo {pages} {010501(R)} (\bibinfo {year} {2009})}\BibitemShut {NoStop}%
\bibitem [{\citenamefont {Ohashi}\ \emph {et~al.}(2011)\citenamefont {Ohashi}, \citenamefont {Yatsurugi}, \citenamefont {Sakaue},\ and\ \citenamefont {Nakamura}}]{Ohashi_2022}%
  \BibitemOpen
  \bibfield  {author} {\bibinfo {author} {\bibfnamefont {H.}~\bibnamefont {Ohashi}}, \bibinfo {author} {\bibfnamefont {J.}~\bibnamefont {Yatsurugi}}, \bibinfo {author} {\bibfnamefont {H.~A.}\ \bibnamefont {Sakaue}},\ and\ \bibinfo {author} {\bibfnamefont {N.}~\bibnamefont {Nakamura}},\ }\bibfield  {title} {\bibinfo {title} {High resolution extreme ultraviolet spectrometer for an electron beam ion trap},\ }\href {https://doi.org/10.1063/1.3618686} {\bibfield  {journal} {\bibinfo  {journal} {Rev. Sci. Instrum.}\ }\textbf {\bibinfo {volume} {82}},\ \bibinfo {pages} {083103} (\bibinfo {year} {2011})}\BibitemShut {NoStop}%
\bibitem [{\citenamefont {Koike}\ \emph {et~al.}(2022)\citenamefont {Koike}, \citenamefont {Suzuki}, \citenamefont {Murakami}, \citenamefont {Kato}, \citenamefont {Tamura},\ and\ \citenamefont {Nakamura}}]{Koike2022}%
  \BibitemOpen
  \bibfield  {author} {\bibinfo {author} {\bibfnamefont {F.}~\bibnamefont {Koike}}, \bibinfo {author} {\bibfnamefont {C.}~\bibnamefont {Suzuki}}, \bibinfo {author} {\bibfnamefont {I.}~\bibnamefont {Murakami}}, \bibinfo {author} {\bibfnamefont {D.}~\bibnamefont {Kato}}, \bibinfo {author} {\bibfnamefont {N.}~\bibnamefont {Tamura}},\ and\ \bibinfo {author} {\bibfnamefont {N.}~\bibnamefont {Nakamura}},\ }\bibfield  {title} {\bibinfo {title} {Z-dependent crossing of excited-state energy levels in highly charged galliumlike lanthanide atomic ions},\ }\href {https://doi.org/10.1103/PhysRevA.105.032802} {\bibfield  {journal} {\bibinfo  {journal} {Phys. Rev. A}\ }\textbf {\bibinfo {volume} {105}},\ \bibinfo {pages} {032802} (\bibinfo {year} {2022})}\BibitemShut {NoStop}%
\bibitem [{\citenamefont {Kramida}\ \emph {et~al.}(2024)\citenamefont {Kramida}, \citenamefont {{Yu.~Ralchenko}}, \citenamefont {Reader},\ and\ \citenamefont {{NIST ASD Team}}}]{kramida_atomic_2023}%
  \BibitemOpen
  \bibfield  {author} {\bibinfo {author} {\bibfnamefont {A.}~\bibnamefont {Kramida}}, \bibinfo {author} {\bibnamefont {{Yu.~Ralchenko}}}, \bibinfo {author} {\bibfnamefont {J.}~\bibnamefont {Reader}},\ and\ \bibinfo {author} {\bibnamefont {{NIST ASD Team}}},\ }\href@noop {} {}\bibinfo {howpublished} {{NIST Atomic Spectra Database (ver. 5.12), [Online]. Available: {\tt{https://physics.nist.gov/asd}} [2025, November 23]. National Institute of Standards and Technology, Gaithersburg, MD.}} (\bibinfo {year} {2024})\BibitemShut {NoStop}%
\bibitem [{\citenamefont {Berglund}\ and\ \citenamefont {Wieser}(2011)}]{Berglund2011}%
  \BibitemOpen
  \bibfield  {author} {\bibinfo {author} {\bibfnamefont {M.}~\bibnamefont {Berglund}}\ and\ \bibinfo {author} {\bibfnamefont {M.~E.}\ \bibnamefont {Wieser}},\ }\bibfield  {title} {\bibinfo {title} {Isotopic compositions of the elements 2009 ({IUPAC} technical report)},\ }\href {https://doi.org/doi:10.1351/PAC-REP-10-06-02} {\bibfield  {journal} {\bibinfo  {journal} {Pure Appl. Chem.}\ }\textbf {\bibinfo {volume} {83}},\ \bibinfo {pages} {397} (\bibinfo {year} {2011})}\BibitemShut {NoStop}%
\bibitem [{\citenamefont {Bradski}(2000)}]{opencv_library}%
  \BibitemOpen
  \bibfield  {author} {\bibinfo {author} {\bibfnamefont {G.}~\bibnamefont {Bradski}},\ }\bibfield  {title} {\bibinfo {title} {{The OpenCV Library}},\ }\href {https://github.com/opencv/opencv} {\bibfield  {journal} {\bibinfo  {journal} {Dr. Dobb's Journal of Software Tools}\ } (\bibinfo {year} {2000})}\BibitemShut {NoStop}%
\bibitem [{\citenamefont {Duncan}(1955)}]{duncan_multiple_1955}%
  \BibitemOpen
  \bibfield  {author} {\bibinfo {author} {\bibfnamefont {D.~B.}\ \bibnamefont {Duncan}},\ }\bibfield  {title} {\bibinfo {title} {Multiple range and multiple f tests},\ }\href {http://www.jstor.org/stable/3001478} {\bibfield  {journal} {\bibinfo  {journal} {Biometrics}\ }\textbf {\bibinfo {volume} {11}},\ \bibinfo {pages} {1} (\bibinfo {year} {1955})}\BibitemShut {NoStop}%
\bibitem [{\citenamefont {Birge}(1932)}]{birge_calculation_1932}%
  \BibitemOpen
  \bibfield  {author} {\bibinfo {author} {\bibfnamefont {R.~T.}\ \bibnamefont {Birge}},\ }\bibfield  {title} {\bibinfo {title} {The calculation of errors by the method of least squares},\ }\href {https://doi.org/10.1103/PhysRev.40.207} {\bibfield  {journal} {\bibinfo  {journal} {Phys. Rev.}\ }\textbf {\bibinfo {volume} {40}},\ \bibinfo {pages} {207} (\bibinfo {year} {1932})}\BibitemShut {NoStop}%
\bibitem [{\citenamefont {Blundell}(1993)}]{blundell_calculations_1993}%
  \BibitemOpen
  \bibfield  {author} {\bibinfo {author} {\bibfnamefont {S.~A.}\ \bibnamefont {Blundell}},\ }\bibfield  {title} {\bibinfo {title} {Calculations of the screened self-energy and vacuum polarization in {Li}-like, {Na}-like, and {Cu}-like ions},\ }\href {https://doi.org/10.1103/PhysRevA.47.1790} {\bibfield  {journal} {\bibinfo  {journal} {Phys. Rev. A}\ }\textbf {\bibinfo {volume} {47}},\ \bibinfo {pages} {1790} (\bibinfo {year} {1993})}\BibitemShut {NoStop}%
\bibitem [{\citenamefont {Gillaspy}\ \emph {et~al.}(2013)\citenamefont {Gillaspy}, \citenamefont {Osin}, \citenamefont {{{Yu.}~Ralchenko}}, \citenamefont {Reader},\ and\ \citenamefont {Blundell}}]{Gillaspy2013}%
  \BibitemOpen
  \bibfield  {author} {\bibinfo {author} {\bibfnamefont {J.~D.}\ \bibnamefont {Gillaspy}}, \bibinfo {author} {\bibfnamefont {D.}~\bibnamefont {Osin}}, \bibinfo {author} {\bibnamefont {{{Yu.}~Ralchenko}}}, \bibinfo {author} {\bibfnamefont {J.}~\bibnamefont {Reader}},\ and\ \bibinfo {author} {\bibfnamefont {S.~A.}\ \bibnamefont {Blundell}},\ }\bibfield  {title} {\bibinfo {title} {{Transition energies of the D lines in Na-like ions}},\ }\href {https://doi.org/10.1103/PhysRevA.87.062503} {\bibfield  {journal} {\bibinfo  {journal} {Phys. Rev. A}\ }\textbf {\bibinfo {volume} {87}},\ \bibinfo {pages} {062503} (\bibinfo {year} {2013})}\BibitemShut {NoStop}%
\bibitem [{\citenamefont {Johnson}\ \emph {et~al.}(1988)\citenamefont {Johnson}, \citenamefont {Blundell},\ and\ \citenamefont {Sapirstein}}]{johnson_many-body_1988-2}%
  \BibitemOpen
  \bibfield  {author} {\bibinfo {author} {\bibfnamefont {W.~R.}\ \bibnamefont {Johnson}}, \bibinfo {author} {\bibfnamefont {S.~A.}\ \bibnamefont {Blundell}},\ and\ \bibinfo {author} {\bibfnamefont {J.}~\bibnamefont {Sapirstein}},\ }\bibfield  {title} {\bibinfo {title} {Many-body perturbation-theory calculations of energy levels along the sodium isoelectronic sequence},\ }\href {https://doi.org/10.1103/PhysRevA.38.2699} {\bibfield  {journal} {\bibinfo  {journal} {Phys. Rev. A}\ }\textbf {\bibinfo {volume} {38}},\ \bibinfo {pages} {2699} (\bibinfo {year} {1988})}\BibitemShut {NoStop}%
\bibitem [{\citenamefont {Staiger}\ \emph {et~al.}(2025)\citenamefont {Staiger}, \citenamefont {Mondeel}, \citenamefont {Blundell}, \citenamefont {Dipti}, \citenamefont {O'Neil}, \citenamefont {Silwal}, \citenamefont {Lapierre}, \citenamefont {Gwinner}, \citenamefont {Tan}, \citenamefont {Gillaspy}, \citenamefont {{{Yu.}~Ralchenko}},\ and\ \citenamefont {Takacs}}]{staiger_measurement_2025}%
  \BibitemOpen
  \bibfield  {author} {\bibinfo {author} {\bibfnamefont {H.}~\bibnamefont {Staiger}}, \bibinfo {author} {\bibfnamefont {G.}~\bibnamefont {Mondeel}}, \bibinfo {author} {\bibfnamefont {S.~A.}\ \bibnamefont {Blundell}}, \bibinfo {author} {\bibnamefont {Dipti}}, \bibinfo {author} {\bibfnamefont {G.}~\bibnamefont {O'Neil}}, \bibinfo {author} {\bibfnamefont {R.}~\bibnamefont {Silwal}}, \bibinfo {author} {\bibfnamefont {A.}~\bibnamefont {Lapierre}}, \bibinfo {author} {\bibfnamefont {G.}~\bibnamefont {Gwinner}}, \bibinfo {author} {\bibfnamefont {J.~N.}\ \bibnamefont {Tan}}, \bibinfo {author} {\bibfnamefont {J.~D.}\ \bibnamefont {Gillaspy}}, \bibinfo {author} {\bibnamefont {{{Yu.}~Ralchenko}}},\ and\ \bibinfo {author} {\bibfnamefont {E.}~\bibnamefont {Takacs}},\ }\bibfield  {title} {\bibinfo {title} {Measurement of {$D$}-line energies in sodiumlike {Ir}},\ }\href {https://doi.org/10.1103/xf52-d6s6} {\bibfield  {journal} {\bibinfo  {journal} {Phys. Rev. A}\ }\textbf {\bibinfo {volume} {112}},\ \bibinfo {pages} {012807}
  (\bibinfo {year} {2025})}\BibitemShut {NoStop}%
\bibitem [{\citenamefont {Sapirstein}\ and\ \citenamefont {Cheng}(2015)}]{sapirstein_s_2015}%
  \BibitemOpen
  \bibfield  {author} {\bibinfo {author} {\bibfnamefont {J.}~\bibnamefont {Sapirstein}}\ and\ \bibinfo {author} {\bibfnamefont {K.~T.}\ \bibnamefont {Cheng}},\ }\bibfield  {title} {\bibinfo {title} {{S}-matrix calculations of energy levels of sodiumlike ions},\ }\href {https://doi.org/10.1103/PhysRevA.91.062508} {\bibfield  {journal} {\bibinfo  {journal} {Phys. Rev. A}\ }\textbf {\bibinfo {volume} {91}},\ \bibinfo {pages} {062508} (\bibinfo {year} {2015})}\BibitemShut {NoStop}%
\bibitem [{\citenamefont {Blundell}(2009)}]{blundell_calculation_2009}%
  \BibitemOpen
  \bibfield  {author} {\bibinfo {author} {\bibfnamefont {S.~A.}\ \bibnamefont {Blundell}},\ }\bibfield  {title} {\bibinfo {title} {Calculation of {QED} corrections in highly charged {Zn}-like ions},\ }\href {https://doi.org/10.1139/p08-065} {\bibfield  {journal} {\bibinfo  {journal} {Can. J. Phys.}\ }\textbf {\bibinfo {volume} {87}},\ \bibinfo {pages} {55} (\bibinfo {year} {2009})}\BibitemShut {NoStop}%
\bibitem [{\citenamefont {Silwal}\ \emph {et~al.}(2022)\citenamefont {Silwal}, \citenamefont {{Dipti}}, \citenamefont {Takacs}, \citenamefont {Dreiling}, \citenamefont {Sanders}, \citenamefont {Gall}, \citenamefont {Rudramadevi}, \citenamefont {Gillaspy},\ and\ \citenamefont {Ralchenko}}]{silwal_spectroscopic_2022}%
  \BibitemOpen
  \bibfield  {author} {\bibinfo {author} {\bibfnamefont {R.}~\bibnamefont {Silwal}}, \bibinfo {author} {\bibnamefont {{Dipti}}}, \bibinfo {author} {\bibfnamefont {E.}~\bibnamefont {Takacs}}, \bibinfo {author} {\bibfnamefont {J.~M.}\ \bibnamefont {Dreiling}}, \bibinfo {author} {\bibfnamefont {S.~C.}\ \bibnamefont {Sanders}}, \bibinfo {author} {\bibfnamefont {A.~C.}\ \bibnamefont {Gall}}, \bibinfo {author} {\bibfnamefont {B.~H.}\ \bibnamefont {Rudramadevi}}, \bibinfo {author} {\bibfnamefont {J.~D.}\ \bibnamefont {Gillaspy}},\ and\ \bibinfo {author} {\bibfnamefont {Y.}~\bibnamefont {Ralchenko}},\ }\bibfield  {title} {\bibinfo {title} {Spectroscopic analysis of {{M-}} and {{N-intrashell}} transitions in {{Co-like}} to {{Na-like Yb}} ions},\ }\href {https://doi.org/10.1088/1361-6455/ac44e1} {\bibfield  {journal} {\bibinfo  {journal} {J. Phys. B}\ }\textbf {\bibinfo {volume} {54}},\ \bibinfo {pages} {245001} (\bibinfo {year} {2022})}\BibitemShut {NoStop}%
\bibitem [{\citenamefont {Zubova}\ \emph {et~al.}(2014)\citenamefont {Zubova}, \citenamefont {Kozhedub}, \citenamefont {Shabaev}, \citenamefont {Tupitsyn}, \citenamefont {Volotka}, \citenamefont {Plunien}, \citenamefont {Brandau},\ and\ \citenamefont {{Stöhlker}}}]{zubova-14}%
  \BibitemOpen
  \bibfield  {author} {\bibinfo {author} {\bibfnamefont {N.~A.}\ \bibnamefont {Zubova}}, \bibinfo {author} {\bibfnamefont {Y.~S.}\ \bibnamefont {Kozhedub}}, \bibinfo {author} {\bibfnamefont {V.~M.}\ \bibnamefont {Shabaev}}, \bibinfo {author} {\bibfnamefont {I.~I.}\ \bibnamefont {Tupitsyn}}, \bibinfo {author} {\bibfnamefont {A.~V.}\ \bibnamefont {Volotka}}, \bibinfo {author} {\bibfnamefont {G.}~\bibnamefont {Plunien}}, \bibinfo {author} {\bibfnamefont {C.}~\bibnamefont {Brandau}},\ and\ \bibinfo {author} {\bibfnamefont {T.}~\bibnamefont {{Stöhlker}}},\ }\bibfield  {title} {\bibinfo {title} {Relativistic calculations of the isotope shifts in highly charged {L}i-like ions},\ }\href {https://doi.org/10.1103/PhysRevA.90.062512} {\bibfield  {journal} {\bibinfo  {journal} {Phys. Rev. A}\ }\textbf {\bibinfo {volume} {90}},\ \bibinfo {pages} {062512} (\bibinfo {year} {2014})}\BibitemShut {NoStop}%
\bibitem [{\citenamefont {Fricke}\ \emph {et~al.}(1995)\citenamefont {Fricke}, \citenamefont {Bernhardt}, \citenamefont {Heilig}, \citenamefont {Schaller}, \citenamefont {Schellenberg}, \citenamefont {Shera},\ and\ \citenamefont {Dejager}}]{fricke_nuclear_1995}%
  \BibitemOpen
  \bibfield  {author} {\bibinfo {author} {\bibfnamefont {G.}~\bibnamefont {Fricke}}, \bibinfo {author} {\bibfnamefont {C.}~\bibnamefont {Bernhardt}}, \bibinfo {author} {\bibfnamefont {K.}~\bibnamefont {Heilig}}, \bibinfo {author} {\bibfnamefont {L.~A.}\ \bibnamefont {Schaller}}, \bibinfo {author} {\bibfnamefont {L.}~\bibnamefont {Schellenberg}}, \bibinfo {author} {\bibfnamefont {E.~B.}\ \bibnamefont {Shera}},\ and\ \bibinfo {author} {\bibfnamefont {C.~W.}\ \bibnamefont {Dejager}},\ }\bibfield  {title} {\bibinfo {title} {Nuclear ground state charge radii from electromagnetic interactions},\ }\href {https://doi.org/10.1006/adnd.1995.1007} {\bibfield  {journal} {\bibinfo  {journal} {At. Data Nucl. Data Tables}\ }\textbf {\bibinfo {volume} {60}},\ \bibinfo {pages} {177} (\bibinfo {year} {1995})}\BibitemShut {NoStop}%
\bibitem [{\citenamefont {Raman}\ \emph {et~al.}(2001)\citenamefont {Raman}, \citenamefont {Jr.},\ and\ \citenamefont {Tikkanen}}]{Raman2001}%
  \BibitemOpen
  \bibfield  {author} {\bibinfo {author} {\bibfnamefont {S.}~\bibnamefont {Raman}}, \bibinfo {author} {\bibfnamefont {C.~W.~N.}\ \bibnamefont {Jr.}},\ and\ \bibinfo {author} {\bibfnamefont {P.}~\bibnamefont {Tikkanen}},\ }\bibfield  {title} {\bibinfo {title} {{Transition probability from the ground to the first-excited $2^+$ state of even–even nuclides}},\ }\href {https://doi.org/10.1006/adnd.2001.0858} {\bibfield  {journal} {\bibinfo  {journal} {At. Data Nucl. Data Tables}\ }\textbf {\bibinfo {volume} {78}},\ \bibinfo {pages} {1} (\bibinfo {year} {2001})}\BibitemShut {NoStop}%
\bibitem [{\citenamefont {Pritychenko}\ \emph {et~al.}(2016)\citenamefont {Pritychenko}, \citenamefont {Birch}, \citenamefont {Singh},\ and\ \citenamefont {Horoi}}]{Pritychenko2016}%
  \BibitemOpen
  \bibfield  {author} {\bibinfo {author} {\bibfnamefont {B.}~\bibnamefont {Pritychenko}}, \bibinfo {author} {\bibfnamefont {M.}~\bibnamefont {Birch}}, \bibinfo {author} {\bibfnamefont {B.}~\bibnamefont {Singh}},\ and\ \bibinfo {author} {\bibfnamefont {M.}~\bibnamefont {Horoi}},\ }\bibfield  {title} {\bibinfo {title} {{Tables of E2 transition probabilities from the first $2^+$ states in even–even nuclei}},\ }\href {https://doi.org/10.1016/j.adt.2015.10.001} {\bibfield  {journal} {\bibinfo  {journal} {At. Data Nucl. Data Tables}\ }\textbf {\bibinfo {volume} {107}},\ \bibinfo {pages} {1} (\bibinfo {year} {2016})}\BibitemShut {NoStop}%
\bibitem [{\citenamefont {Georg}\ \emph {et~al.}(1998)\citenamefont {Georg}, \citenamefont {Borchers}, \citenamefont {Keim}, \citenamefont {Klein}, \citenamefont {Lievens}, \citenamefont {Neugart}, \citenamefont {Neuroth}, \citenamefont {Rao}, \citenamefont {Schulz},\ and\ \citenamefont {{the {ISOLDE} Collaboration}}}]{GBK98}%
  \BibitemOpen
  \bibfield  {author} {\bibinfo {author} {\bibfnamefont {U.}~\bibnamefont {Georg}}, \bibinfo {author} {\bibfnamefont {W.}~\bibnamefont {Borchers}}, \bibinfo {author} {\bibfnamefont {M.}~\bibnamefont {Keim}}, \bibinfo {author} {\bibfnamefont {A.}~\bibnamefont {Klein}}, \bibinfo {author} {\bibfnamefont {P.}~\bibnamefont {Lievens}}, \bibinfo {author} {\bibfnamefont {R.}~\bibnamefont {Neugart}}, \bibinfo {author} {\bibfnamefont {M.}~\bibnamefont {Neuroth}}, \bibinfo {author} {\bibfnamefont {P.~M.}\ \bibnamefont {Rao}}, \bibinfo {author} {\bibfnamefont {C.}~\bibnamefont {Schulz}},\ and\ \bibinfo {author} {\bibnamefont {{the {ISOLDE} Collaboration}}},\ }\bibfield  {title} {\bibinfo {title} {Laser spectroscopy investigation of the nuclear moments and radii of lutetium isotopes},\ }\href {https://doi.org/https://doi.org/10.1007/s100500050172} {\bibfield  {journal} {\bibinfo  {journal} {Eur. Phys. J. A}\ }\textbf {\bibinfo {volume} {3}},\ \bibinfo {pages} {225} (\bibinfo {year} {1998})}\BibitemShut {NoStop}%
\bibitem [{\citenamefont {Kl{\"{u}}pfel}\ \emph {et~al.}(2009)\citenamefont {Kl{\"{u}}pfel}, \citenamefont {Reinhard}, \citenamefont {B{\"{u}}rvenich},\ and\ \citenamefont {Maruhn}}]{Kluepfel2009}%
  \BibitemOpen
  \bibfield  {author} {\bibinfo {author} {\bibfnamefont {P.}~\bibnamefont {Kl{\"{u}}pfel}}, \bibinfo {author} {\bibfnamefont {P.-G.}\ \bibnamefont {Reinhard}}, \bibinfo {author} {\bibfnamefont {T.~J.}\ \bibnamefont {B{\"{u}}rvenich}},\ and\ \bibinfo {author} {\bibfnamefont {J.~A.}\ \bibnamefont {Maruhn}},\ }\bibfield  {title} {\bibinfo {title} {Variations on a theme by {Skyrme}: A systematic study of adjustments of model parameters},\ }\href {https://doi.org/10.1103/PhysRevC.79.034310} {\bibfield  {journal} {\bibinfo  {journal} {Phys. Rev. C}\ }\textbf {\bibinfo {volume} {79}},\ \bibinfo {pages} {034310} (\bibinfo {year} {2009})}\BibitemShut {NoStop}%
\bibitem [{\citenamefont {{Miller}}\ \emph {et~al.}(2019)\citenamefont {{Miller}}, \citenamefont {{Minamisono}}, \citenamefont {{Klose}}, \citenamefont {{Garand}}, \citenamefont {{Kujawa}}, \citenamefont {{Lantis}}, \citenamefont {{\ Liu}}, \citenamefont {{Maa{\ss}}}, \citenamefont {{Mantica}}, \citenamefont {{Nazarewicz}}, \citenamefont {{N{\"o}rtersh{\"a}user}}, \citenamefont {{Pineda}}, \citenamefont {{Re\ inhard}}, \citenamefont {{Rossi}}, \citenamefont {{Sommer}} \emph {et~al.}}]{Miller2019}%
  \BibitemOpen
  \bibfield  {author} {\bibinfo {author} {\bibfnamefont {A.~J.}\ \bibnamefont {{Miller}}}, \bibinfo {author} {\bibfnamefont {K.}~\bibnamefont {{Minamisono}}}, \bibinfo {author} {\bibfnamefont {A.}~\bibnamefont {{Klose}}}, \bibinfo {author} {\bibfnamefont {D.}~\bibnamefont {{Garand}}}, \bibinfo {author} {\bibfnamefont {C.}~\bibnamefont {{Kujawa}}}, \bibinfo {author} {\bibfnamefont {J.~D.}\ \bibnamefont {{Lantis}}}, \bibinfo {author} {\bibfnamefont {Y.}~\bibnamefont {{\ Liu}}}, \bibinfo {author} {\bibfnamefont {B.}~\bibnamefont {{Maa{\ss}}}}, \bibinfo {author} {\bibfnamefont {P.~F.}\ \bibnamefont {{Mantica}}}, \bibinfo {author} {\bibfnamefont {W.}~\bibnamefont {{Nazarewicz}}}, \bibinfo {author} {\bibfnamefont {W.}~\bibnamefont {{N{\"o}rtersh{\"a}user}}}, \bibinfo {author} {\bibfnamefont {S.~V.}\ \bibnamefont {{Pineda}}}, \bibinfo {author} {\bibfnamefont {P.~G.}\ \bibnamefont {{Re\ inhard}}}, \bibinfo {author} {\bibfnamefont {D.~M.}\ \bibnamefont {{Rossi}}}, \bibinfo {author} {\bibfnamefont {F.}~\bibnamefont
  {{Sommer}}}, \emph {et~al.},\ }\bibfield  {title} {\bibinfo {title} {{Proton superfluidity and charge radii in proton-rich calcium isotopes}},\ }\href {https://doi.org/10.1038/s41567-019-0416-9} {\bibfield  {journal} {\bibinfo  {journal} {Nat. Phys.}\ }\textbf {\bibinfo {volume} {15}},\ \bibinfo {pages} {432} (\bibinfo {year} {2019})}\BibitemShut {NoStop}%
\bibitem [{\citenamefont {Karthein}\ \emph {et~al.}(2023)\citenamefont {Karthein}, \citenamefont {Ricketts}, \citenamefont {Garcia~Ruiz}, \citenamefont {Billowes}, \citenamefont {Binnersley}, \citenamefont {Cocolios}, \citenamefont {Dobaczewski}, \citenamefont {Farooq-Smith}, \citenamefont {Flanagan}, \citenamefont {Georgiev}, \citenamefont {Gins}, \citenamefont {de~Groote}, \citenamefont {Gustafsson}, \citenamefont {Holt} \emph {et~al.}}]{Karthein2024}%
  \BibitemOpen
  \bibfield  {author} {\bibinfo {author} {\bibfnamefont {J.}~\bibnamefont {Karthein}}, \bibinfo {author} {\bibfnamefont {C.}~\bibnamefont {Ricketts}}, \bibinfo {author} {\bibfnamefont {R.}~\bibnamefont {Garcia~Ruiz}}, \bibinfo {author} {\bibfnamefont {J.}~\bibnamefont {Billowes}}, \bibinfo {author} {\bibfnamefont {C.}~\bibnamefont {Binnersley}}, \bibinfo {author} {\bibfnamefont {T.}~\bibnamefont {Cocolios}}, \bibinfo {author} {\bibfnamefont {J.}~\bibnamefont {Dobaczewski}}, \bibinfo {author} {\bibfnamefont {G.}~\bibnamefont {Farooq-Smith}}, \bibinfo {author} {\bibfnamefont {K.}~\bibnamefont {Flanagan}}, \bibinfo {author} {\bibfnamefont {G.}~\bibnamefont {Georgiev}}, \bibinfo {author} {\bibfnamefont {W.}~\bibnamefont {Gins}}, \bibinfo {author} {\bibfnamefont {R.}~\bibnamefont {de~Groote}}, \bibinfo {author} {\bibfnamefont {F.~P.}\ \bibnamefont {Gustafsson}}, \bibinfo {author} {\bibfnamefont {J.}~\bibnamefont {Holt}}, \emph {et~al.},\ }\bibfield  {title} {\bibinfo {title} {Electromagnetic properties of indium
  isotopes elucidate the doubly magic character of \textsuperscript{100}{Sn}},\ }\href@noop {} {\bibfield  {journal} {\bibinfo  {journal} {Nat. Phys.}\ } (\bibinfo {year} {2023})}\BibitemShut {NoStop}%
\bibitem [{\citenamefont {Reinhard}\ and\ \citenamefont {Nazarewicz}(2017)}]{Reinhard2017global}%
  \BibitemOpen
  \bibfield  {author} {\bibinfo {author} {\bibfnamefont {P.-G.}\ \bibnamefont {Reinhard}}\ and\ \bibinfo {author} {\bibfnamefont {W.}~\bibnamefont {Nazarewicz}},\ }\bibfield  {title} {\bibinfo {title} {Toward a global description of nuclear charge radii: Exploring the fayans energy density functional},\ }\href {https://doi.org/10.1103/PhysRevC.95.064328} {\bibfield  {journal} {\bibinfo  {journal} {Phys. Rev. C}\ }\textbf {\bibinfo {volume} {95}},\ \bibinfo {pages} {064328} (\bibinfo {year} {2017})}\BibitemShut {NoStop}%
\bibitem [{\citenamefont {Reinhard}\ \emph {et~al.}(2021)\citenamefont {Reinhard}, \citenamefont {Schuetrumpf},\ and\ \citenamefont {Maruhn}}]{SkyAx2021}%
  \BibitemOpen
  \bibfield  {author} {\bibinfo {author} {\bibfnamefont {P.-G.}\ \bibnamefont {Reinhard}}, \bibinfo {author} {\bibfnamefont {B.}~\bibnamefont {Schuetrumpf}},\ and\ \bibinfo {author} {\bibfnamefont {J.}~\bibnamefont {Maruhn}},\ }\bibfield  {title} {\bibinfo {title} {The axial {H}artree-{F}ock + {BCS} code {S}ky{A}x},\ }\href {https://doi.org/10.1016/j.cpc.2020.107603} {\bibfield  {journal} {\bibinfo  {journal} {Comput. Phys. Commun.}\ }\textbf {\bibinfo {volume} {258}},\ \bibinfo {pages} {107603} (\bibinfo {year} {2021})}\BibitemShut {NoStop}%
\bibitem [{\citenamefont {Dobaczewski}\ \emph {et~al.}(2014)\citenamefont {Dobaczewski}, \citenamefont {Nazarewicz},\ and\ \citenamefont {Reinhard}}]{Dob14}%
  \BibitemOpen
  \bibfield  {author} {\bibinfo {author} {\bibfnamefont {J.}~\bibnamefont {Dobaczewski}}, \bibinfo {author} {\bibfnamefont {W.}~\bibnamefont {Nazarewicz}},\ and\ \bibinfo {author} {\bibfnamefont {P.-G.}\ \bibnamefont {Reinhard}},\ }\bibfield  {title} {\bibinfo {title} {Error estimates of theoretical models: a guide},\ }\href {https://doi.org/10.1088/0954-3899/41/7/074001} {\bibfield  {journal} {\bibinfo  {journal} {J. Phys. G}\ }\textbf {\bibinfo {volume} {41}},\ \bibinfo {pages} {074001} (\bibinfo {year} {2014})}\BibitemShut {NoStop}%
\bibitem [{\citenamefont {Ring}\ and\ \citenamefont {Schuck}(1980)}]{(Rin80b)}%
  \BibitemOpen
  \bibfield  {author} {\bibinfo {author} {\bibfnamefont {P.}~\bibnamefont {Ring}}\ and\ \bibinfo {author} {\bibfnamefont {P.}~\bibnamefont {Schuck}},\ }\href {https://www.springer.com/gp/book/9783540212065} {\emph {\bibinfo {title} {The Nuclear Many-Body Problem}}}\ (\bibinfo  {publisher} {Springer-Verlag, Berlin},\ \bibinfo {year} {1980})\BibitemShut {NoStop}%
\bibitem [{\citenamefont {Nazarewicz}\ \emph {et~al.}(1990)\citenamefont {Nazarewicz}, \citenamefont {Riley},\ and\ \citenamefont {Garrett}}]{Nazarewicz1990}%
  \BibitemOpen
  \bibfield  {author} {\bibinfo {author} {\bibfnamefont {W.}~\bibnamefont {Nazarewicz}}, \bibinfo {author} {\bibfnamefont {M.}~\bibnamefont {Riley}},\ and\ \bibinfo {author} {\bibfnamefont {J.}~\bibnamefont {Garrett}},\ }\bibfield  {title} {\bibinfo {title} {Equilibrium deformations and excitation energies of single-quasiproton band heads of rare-earth nuclei},\ }\href {https://doi.org/10.1016/0375-9474(90)90004-6} {\bibfield  {journal} {\bibinfo  {journal} {Nucl. Phys. A}\ }\textbf {\bibinfo {volume} {512}},\ \bibinfo {pages} {61} (\bibinfo {year} {1990})}\BibitemShut {NoStop}%
\bibitem [{\citenamefont {Reinhard}\ and\ \citenamefont {Nazarewicz}(2021)}]{DFTformfactors}%
  \BibitemOpen
  \bibfield  {author} {\bibinfo {author} {\bibfnamefont {P.-G.}\ \bibnamefont {Reinhard}}\ and\ \bibinfo {author} {\bibfnamefont {W.}~\bibnamefont {Nazarewicz}},\ }\bibfield  {title} {\bibinfo {title} {Nuclear charge densities in spherical and deformed nuclei: Toward precise calculations of charge radii},\ }\href {https://doi.org/10.1103/PhysRevC.103.054310} {\bibfield  {journal} {\bibinfo  {journal} {Phys. Rev. C}\ }\textbf {\bibinfo {volume} {103}},\ \bibinfo {pages} {054310} (\bibinfo {year} {2021})}\BibitemShut {NoStop}%
\bibitem [{\citenamefont {Reinhard}\ and\ \citenamefont {Otten}(1984)}]{Reinhard1984}%
  \BibitemOpen
  \bibfield  {author} {\bibinfo {author} {\bibfnamefont {P.-G.}\ \bibnamefont {Reinhard}}\ and\ \bibinfo {author} {\bibfnamefont {E.~W.}\ \bibnamefont {Otten}},\ }\bibfield  {title} {\bibinfo {title} {Transition to deformed shapes as a nuclear {Jahn-Teller} effect},\ }\href {https://doi.org/10.1016/0375-9474(84)90437-8} {\bibfield  {journal} {\bibinfo  {journal} {Nucl. Phys. A}\ }\textbf {\bibinfo {volume} {420}},\ \bibinfo {pages} {173} (\bibinfo {year} {1984})}\BibitemShut {NoStop}%
\bibitem [{\citenamefont {Nazarewicz}(1994)}]{Nazarewicz1994}%
  \BibitemOpen
  \bibfield  {author} {\bibinfo {author} {\bibfnamefont {W.}~\bibnamefont {Nazarewicz}},\ }\bibfield  {title} {\bibinfo {title} {Microscopic origin of nuclear deformations},\ }\href {https://doi.org/10.1016/0375-9474(94)90037-X} {\bibfield  {journal} {\bibinfo  {journal} {Nucl. Phys. A}\ }\textbf {\bibinfo {volume} {574}},\ \bibinfo {pages} {27 } (\bibinfo {year} {1994})}\BibitemShut {NoStop}%
\bibitem [{\citenamefont {Reinhard}\ and\ \citenamefont {Nazarewicz}(2022)}]{Reinhard2022}%
  \BibitemOpen
  \bibfield  {author} {\bibinfo {author} {\bibfnamefont {P.-G.}\ \bibnamefont {Reinhard}}\ and\ \bibinfo {author} {\bibfnamefont {W.}~\bibnamefont {Nazarewicz}},\ }\bibfield  {title} {\bibinfo {title} {Statistical correlations of nuclear quadrupole deformations and charge radii},\ }\href {https://doi.org/10.1103/PhysRevC.106.014303} {\bibfield  {journal} {\bibinfo  {journal} {Phys. Rev. C}\ }\textbf {\bibinfo {volume} {106}},\ \bibinfo {pages} {014303} (\bibinfo {year} {2022})}\BibitemShut {NoStop}%
\bibitem [{\citenamefont {Barzakh}\ \emph {et~al.}(1998)\citenamefont {Barzakh}, \citenamefont {Chubukov}, \citenamefont {Fedorov}, \citenamefont {Moroz}, \citenamefont {Panteleev}, \citenamefont {Seliverstov},\ and\ \citenamefont {Volkov}}]{Barzakh1998}%
  \BibitemOpen
  \bibfield  {author} {\bibinfo {author} {\bibfnamefont {A.~E.}\ \bibnamefont {Barzakh}}, \bibinfo {author} {\bibfnamefont {I.~Y.}\ \bibnamefont {Chubukov}}, \bibinfo {author} {\bibfnamefont {D.~V.}\ \bibnamefont {Fedorov}}, \bibinfo {author} {\bibfnamefont {F.~V.}\ \bibnamefont {Moroz}}, \bibinfo {author} {\bibfnamefont {V.~N.}\ \bibnamefont {Panteleev}}, \bibinfo {author} {\bibfnamefont {M.~D.}\ \bibnamefont {Seliverstov}},\ and\ \bibinfo {author} {\bibfnamefont {Y.~M.}\ \bibnamefont {Volkov}},\ }\bibfield  {title} {\bibinfo {title} {Isotope shift and hyperfine structure measurements for \textsuperscript{155}{Yb} by laser ion source technique},\ }\href {https://doi.org/10.1007/s100500050023} {\bibfield  {journal} {\bibinfo  {journal} {The European Physical Journal A}\ }\textbf {\bibinfo {volume} {1}},\ \bibinfo {pages} {3} (\bibinfo {year} {1998})}\BibitemShut {NoStop}%
\bibitem [{\citenamefont {Barzakh}\ \emph {et~al.}(2000)\citenamefont {Barzakh}, \citenamefont {Chubukov}, \citenamefont {Fedorov}, \citenamefont {Panteleev}, \citenamefont {Seliverstov},\ and\ \citenamefont {Volkov}}]{Barzakh2000}%
  \BibitemOpen
  \bibfield  {author} {\bibinfo {author} {\bibfnamefont {A.~E.}\ \bibnamefont {Barzakh}}, \bibinfo {author} {\bibfnamefont {I.~Y.}\ \bibnamefont {Chubukov}}, \bibinfo {author} {\bibfnamefont {D.~V.}\ \bibnamefont {Fedorov}}, \bibinfo {author} {\bibfnamefont {V.~N.}\ \bibnamefont {Panteleev}}, \bibinfo {author} {\bibfnamefont {M.~D.}\ \bibnamefont {Seliverstov}},\ and\ \bibinfo {author} {\bibfnamefont {Y.~M.}\ \bibnamefont {Volkov}},\ }\bibfield  {title} {\bibinfo {title} {Mean square charge radii of the neutron-deficient rare-earth isotopes in the region of the nuclear shell {$N\approx 82$} measured by the laser ion source spectroscopy technique},\ }\href {https://doi.org/10.1103/PhysRevC.61.034304} {\bibfield  {journal} {\bibinfo  {journal} {Phys. Rev. C}\ }\textbf {\bibinfo {volume} {61}},\ \bibinfo {pages} {034304} (\bibinfo {year} {2000})}\BibitemShut {NoStop}%
\bibitem [{\citenamefont {Schulz}\ \emph {et~al.}(1991)\citenamefont {Schulz}, \citenamefont {Arnold}, \citenamefont {Borchers}, \citenamefont {Neu}, \citenamefont {Neugart}, \citenamefont {Neuroth}, \citenamefont {Otten}, \citenamefont {Scherf}, \citenamefont {Wendt}, \citenamefont {Lievens}, \citenamefont {Kudryavtsev}, \citenamefont {Letokhov}, \citenamefont {Mishin}, \citenamefont {Petrunin},\ and\ \citenamefont {the ISOLDE~Collaboration}}]{Schulz1991}%
  \BibitemOpen
  \bibfield  {author} {\bibinfo {author} {\bibfnamefont {C.}~\bibnamefont {Schulz}}, \bibinfo {author} {\bibfnamefont {E.}~\bibnamefont {Arnold}}, \bibinfo {author} {\bibfnamefont {W.}~\bibnamefont {Borchers}}, \bibinfo {author} {\bibfnamefont {W.}~\bibnamefont {Neu}}, \bibinfo {author} {\bibfnamefont {R.}~\bibnamefont {Neugart}}, \bibinfo {author} {\bibfnamefont {M.}~\bibnamefont {Neuroth}}, \bibinfo {author} {\bibfnamefont {E.~W.}\ \bibnamefont {Otten}}, \bibinfo {author} {\bibfnamefont {M.}~\bibnamefont {Scherf}}, \bibinfo {author} {\bibfnamefont {K.}~\bibnamefont {Wendt}}, \bibinfo {author} {\bibfnamefont {P.}~\bibnamefont {Lievens}}, \bibinfo {author} {\bibfnamefont {Y.~A.}\ \bibnamefont {Kudryavtsev}}, \bibinfo {author} {\bibfnamefont {V.~S.}\ \bibnamefont {Letokhov}}, \bibinfo {author} {\bibfnamefont {V.~I.}\ \bibnamefont {Mishin}}, \bibinfo {author} {\bibfnamefont {V.~V.}\ \bibnamefont {Petrunin}},\ and\ \bibinfo {author} {\bibnamefont {the ISOLDE~Collaboration}},\ }\bibfield  {title} {\bibinfo
  {title} {Resonance ionization spectroscopy on a fast atomic ytterbium beam},\ }\href {https://doi.org/10.1088/0953-4075/24/22/020} {\bibfield  {journal} {\bibinfo  {journal} {J. Phys. B}\ }\textbf {\bibinfo {volume} {24}},\ \bibinfo {pages} {4831} (\bibinfo {year} {1991})}\BibitemShut {NoStop}%
\bibitem [{\citenamefont {Jin}\ \emph {et~al.}(1991)\citenamefont {Jin}, \citenamefont {Horiguchi}, \citenamefont {Wakasugi}, \citenamefont {Hasegawa},\ and\ \citenamefont {Yang}}]{Jin1991}%
  \BibitemOpen
  \bibfield  {author} {\bibinfo {author} {\bibfnamefont {W.-G.}\ \bibnamefont {Jin}}, \bibinfo {author} {\bibfnamefont {T.}~\bibnamefont {Horiguchi}}, \bibinfo {author} {\bibfnamefont {M.}~\bibnamefont {Wakasugi}}, \bibinfo {author} {\bibfnamefont {T.}~\bibnamefont {Hasegawa}},\ and\ \bibinfo {author} {\bibfnamefont {W.}~\bibnamefont {Yang}},\ }\bibfield  {title} {\bibinfo {title} {Systematic study of isotope shifts and hyperfine structures in {Yb} {I} by atomic-beam laser spectroscopy},\ }\href {https://doi.org/10.1143/JPSJ.60.2896} {\bibfield  {journal} {\bibinfo  {journal} {Journal of the Physical Society of Japan}\ }\textbf {\bibinfo {volume} {60}},\ \bibinfo {pages} {2896} (\bibinfo {year} {1991})}\BibitemShut {NoStop}%
\bibitem [{\citenamefont {Sprouse}\ \emph {et~al.}(1989)\citenamefont {Sprouse}, \citenamefont {Das}, \citenamefont {Lauritsen}, \citenamefont {Schecker}, \citenamefont {Berger}, \citenamefont {Billowes}, \citenamefont {Holbrow}, \citenamefont {Mahnke},\ and\ \citenamefont {Rolston}}]{Sprouse1989}%
  \BibitemOpen
  \bibfield  {author} {\bibinfo {author} {\bibfnamefont {G.~D.}\ \bibnamefont {Sprouse}}, \bibinfo {author} {\bibfnamefont {J.}~\bibnamefont {Das}}, \bibinfo {author} {\bibfnamefont {T.}~\bibnamefont {Lauritsen}}, \bibinfo {author} {\bibfnamefont {J.}~\bibnamefont {Schecker}}, \bibinfo {author} {\bibfnamefont {A.}~\bibnamefont {Berger}}, \bibinfo {author} {\bibfnamefont {J.}~\bibnamefont {Billowes}}, \bibinfo {author} {\bibfnamefont {C.~H.}\ \bibnamefont {Holbrow}}, \bibinfo {author} {\bibfnamefont {H.-E.}\ \bibnamefont {Mahnke}},\ and\ \bibinfo {author} {\bibfnamefont {S.~L.}\ \bibnamefont {Rolston}},\ }\bibfield  {title} {\bibinfo {title} {Laser spectroscopy of light {Yb} isotopes on-line in a cooled gas cell},\ }\href {https://doi.org/10.1103/PhysRevLett.63.1463} {\bibfield  {journal} {\bibinfo  {journal} {Phys. Rev. Lett.}\ }\textbf {\bibinfo {volume} {63}},\ \bibinfo {pages} {1463} (\bibinfo {year} {1989})}\BibitemShut {NoStop}%
\bibitem [{\citenamefont {Flanagan}\ \emph {et~al.}(2025)\citenamefont {Flanagan}, \citenamefont {Staiger}, \citenamefont {Takacs},\ and\ \citenamefont {Dimitrou}}]{iaea_summary_2025}%
  \BibitemOpen
  \bibfield  {author} {\bibinfo {author} {\bibfnamefont {K.}~\bibnamefont {Flanagan}}, \bibinfo {author} {\bibfnamefont {H.}~\bibnamefont {Staiger}}, \bibinfo {author} {\bibfnamefont {E.}~\bibnamefont {Takacs}},\ and\ \bibinfo {author} {\bibfnamefont {P.}~\bibnamefont {Dimitrou}},\ }\href {https://doi.org/10.61092/iaea.vm5h-hmep} {\bibinfo {title} {Compilation and {Evaluation of Nuclear Charge Radii} - {Summary Report of the Technical Meeting}}} (\bibinfo {year} {2025})\BibitemShut {NoStop}%
\bibitem [{\citenamefont {Kawasaki}\ \emph {et~al.}(2024)\citenamefont {Kawasaki}, \citenamefont {Kobayashi}, \citenamefont {Nishiyama}, \citenamefont {Tanabe},\ and\ \citenamefont {Yasuda}}]{kawasaki_isotopeshift_2024}%
  \BibitemOpen
  \bibfield  {author} {\bibinfo {author} {\bibfnamefont {A.}~\bibnamefont {Kawasaki}}, \bibinfo {author} {\bibfnamefont {T.}~\bibnamefont {Kobayashi}}, \bibinfo {author} {\bibfnamefont {A.}~\bibnamefont {Nishiyama}}, \bibinfo {author} {\bibfnamefont {T.}~\bibnamefont {Tanabe}},\ and\ \bibinfo {author} {\bibfnamefont {M.}~\bibnamefont {Yasuda}},\ }\bibfield  {title} {\bibinfo {title} {Isotope-shift analysis with the $4f^{14}6s^2\ ^1s_0$ -- $4f^{13}5d6s^2\ (j=2)$ transition in ytterbium},\ }\href {https://doi.org/10.1103/PhysRevA.109.062806} {\bibfield  {journal} {\bibinfo  {journal} {Phys. Rev. A}\ }\textbf {\bibinfo {volume} {109}},\ \bibinfo {pages} {062806} (\bibinfo {year} {2024})}\BibitemShut {NoStop}%
\bibitem [{\citenamefont {Kariya}\ and\ \citenamefont {Kurata}(2004)}]{kariya_generalized_2004}%
  \BibitemOpen
  \bibfield  {author} {\bibinfo {author} {\bibfnamefont {T.}~\bibnamefont {Kariya}}\ and\ \bibinfo {author} {\bibfnamefont {H.}~\bibnamefont {Kurata}},\ }\href {https://doi.org/10.1002/0470866993} {\emph {\bibinfo {title} {Generalized {{Least Squares}}}}}\ (\bibinfo  {publisher} {John Wiley \& Sons},\ \bibinfo {year} {2004})\BibitemShut {NoStop}%
\end{thebibliography}%

\end{document}